\documentclass[final,3p,times,authoryear]{elsarticle}

\usepackage{xcolor}
\usepackage{amsmath}
\usepackage{natbib}
\biboptions{authoryear,round}
\usepackage{mathtools}
\usepackage{physics}
\usepackage{stmaryrd} 
\usepackage{epstopdf}
\usepackage{bm}
\usepackage{graphicx,wrapfig}
\usepackage[T1]{fontenc}
\usepackage[normalem]{ulem}
\usepackage{algorithm}
\usepackage{algpseudocode}

\newcommand{\stkout}[1]{\ifmmode\text{\sout{\ensuremath{#1}}}\else\sout{#1}\fi}
\usepackage{mathrsfs}  
\usepackage{amssymb}
\usepackage{nomencl}
\usepackage{etoolbox}
\usepackage{multicol}
\usepackage{caption,subcaption}
\usepackage{tabularx}
\usepackage{booktabs}
\usepackage{enumitem}
\usepackage[title]{appendix}
\usepackage{tikz}
\usetikzlibrary{arrows.meta,positioning,calc,decorations.pathreplacing,shapes.geometric}
\newtheorem{remark}{Remark}%
\makenomenclature

\begin{document}

\begin{frontmatter}




\title{On the complementary roles of anisotropic crack density and anisotropic crack driving force in phase-field modeling of mixed-mode fracture}

\author[1]{Guk Heon Kim\fnref{fn_equal}}
\author[1]{Minseo Kim\fnref{fn_equal}}
\fntext[fn_equal]{These authors contributed equally to this work.}
\author[2]{Kwangsan Chun\corref{cor2}}
\cortext[cor2]{Corresponding Author}
\ead{kschun@mnu.ac.kr}
\author[1]{Jaemin Kim\corref{cor1}}
\cortext[cor1]{Corresponding Author}
\ead{jaeminkim@changwon.ac.kr}

\address[1]{School of Mechanical Engineering, Changwon National University, Changwon 51140, Republic of Korea}
\address[2]{Department of Smart Shipbuilding Systems, Mokpo National University, Muan, 58554, Republic of Korea

}

\begin{abstract}
Phase-field models for anisotropic fracture employ two complementary mechanisms: (i)~the anisotropic crack density function, controlling direction-dependent fracture resistance, and (ii)~the anisotropic strain energy, governing the fracture driving force. Although the unified framework was presented in Pranavi et al.\ [Comput.\ Mech., 73 (2024)], the distinct roles of these mechanisms and their interaction remain uninvestigated. This work addresses this gap by first validating the formulation against mixed-mode fracture experiments on a soft elastomer (Lu et al.\ [Extreme Mech.\ Lett., 48 (2021)]), and then conducting systematic parametric studies on single-edge-notched (SEN) and open-hole tension (OHT) specimens to isolate each mechanism. The SEN studies show that the crack density anisotropy controls the crack path and toughness while leaving the elastic response unchanged, whereas the anisotropic strain energy deflects the crack but saturates rapidly. The OHT studies reveal a geometry-dependent role expansion: the anisotropic strain energy governs fiber-orientation-dependent stiffness, peak force, and fracture displacement. When both mechanisms act together, the combined response exhibits nonlinear synergistic interaction exceeding the linear sum of the individual contributions. These results establish that the crack density anisotropy governs the crack path (fracture resistance), while the anisotropic strain energy governs the driving force and, in stress-concentration geometries, additionally controls the elastic strain energy distribution around the stress concentrator.
\end{abstract}

\end{frontmatter}

\section*{Nomenclature}
\begin{center}
  \begin{tabularx}{\textwidth}{l@{\hspace{4mm}}X l@{\hspace{4mm}}X}
    \hline
    Symbol & Definition & Symbol & Definition \\
    \hline
    $\mathbf{X},\mathbf{x}$ & Reference / current position & $\mathbf{u}$ & Displacement field \\
    $\boldsymbol{\varphi},\mathbf{F}$ & Deformation mapping / gradient & $J$ & Jacobian determinant \\
    $\mathbf{C},\bar{\mathbf{C}}$ & Right Cauchy--Green tensor / isochoric part& $\lambda_i,\bar{\lambda}_i$ & Principal / isochoric stretches \\
    $\mathbf{P},\boldsymbol{\sigma}$ & 1st Piola--Kirchhoff / Cauchy stress & $\mu,\kappa$ & Shear / bulk modulus \\
    $\Psi,\Psi_{\mathrm{ani}}$ & Helmholtz free energy / anisotropic energy & $\Psi^{+},\Psi^{-}$ & Tensile / compressive energies \\
    $\Psi_{\mathrm{v}},\Psi_{\mathrm{d}}$ & Volumetric / deviatoric energy & $n,b$ & GNH exponent / hardening parameter \\
    $d,\ell_{d}$ & Phase-field damage / length scale & $g(d),\alpha(d)$ & Degradation / geometric crack function \\
    $G_{c}$ & Fracture toughness & $G_{c,\mathrm{I}},G_{c,\mathrm{II}}$ & Mode-I / mode-II toughness \\
    $\gamma_c,c_{w}$ & Crack density / normalization & $\mathcal{H}$ & Normalized driving force \\
    $k_{\ell}$ & Residual stiffness & $\mathbf{a},\theta_f$ & Fiber direction / angle \\
    $\mathbf{A}$ & Structural tensor & $\beta$ & Anisotropy strength \\
    $p,q,r$ & Power-law exponents & $\omega,\boldsymbol{\xi}$ & Scalar / vector microstresses \\
    $\Omega_0$ & Reference domain & $\partial\Omega_{0,u},\partial\Omega_{0,t}$ & Dirichlet / Neumann boundaries \\
    \hline
  \end{tabularx}
\end{center}


\newpage
\section{Introduction}\label{sec:1}
The computational modeling of fracture progresses through several methodological stages. Early remeshing-based approaches require costly mesh updates at each propagation step. The extended/generalized finite element method (XFEM/GFEM) alleviates this by enriching the approximation space with discontinuous and near-tip functions, enabling cracks to propagate independently of the mesh \citep{moes1999finite,belytschko2009review}; however, robust implementation in three dimensions and for multiple interacting cracks remains challenging. Cohesive zone models insert traction--separation relations along predefined element interfaces \citep{xu1994numerical,park2011cohesive}, but their accuracy depends on element alignment with the unknown crack path. Phase-field fracture (PFF) methods emerge as an alternative paradigm that circumvents these difficulties, offering a mathematically robust means to simulate crack nucleation, propagation, branching, and merging in arbitrary geometries without the need for ad hoc tracking algorithms \citep{francfort1998revisiting, bourdin2000numerical, miehe2010thermodynamically, kim2025phase, kim2026unified}. By regularizing the sharp crack interface into a diffuse damage field, the method overcomes the mesh-dependency and topological difficulties associated with discrete fracture mechanics. However, this challenge is even greater when anisotropic fracture resistance and mixed-mode loading are taken into account.

In many engineering materials---such as fiber-reinforced composites, sedimentary rocks, and printed laminates---fracture resistance is highly directional, and failure is driven by a complex combination of tensile and shear modes. Standard isotropic phase-field models, which typically employ a scalar fracture toughness and simplified energy degradation functions, struggle to capture these directional effects. Moreover, distinguishing between the energy contributions of different fracture modes (Mode I opening vs. Mode II shear) within the variational framework often requires spectral decompositions or phenomenological criteria that lack a direct thermodynamic interpretation \citep{wu2020anisotropic, kristensen2021assessment}. Recent anisotropic phase-field models for fiber-reinforced composites make significant progress: \citet{dean2020anisotropic} propose an anisotropic cohesive phase-field model for quasi-brittle intra-laminar fracture, \citet{mandal2020length} develop a length-scale insensitive anisotropic model for hyperelastic composites, and \citet{yu2025unified} introduce a unified anisotropic model with a novel crack surface density function combining features of double and single phase-field formulations. However, these works primarily rely on stress- or strain-based failure criteria (e.g., Hashin type) for mode decomposition rather than an orientation-dependent energy/toughness distribution.

A critical issue in mixed-mode phase-field modeling is the treatment of mode mixity and irreversibility. Many formulations rely on historical maximum variables to enforce the irreversibility condition ($\dot{d} \ge 0$), which effectively decouples the current damage state from the instantaneous energy landscape. Additionally, classical flow rules or stress-based criteria for mixed-mode failure are often difficult to integrate into the energetic phase-field structure without violating variational consistency. Recent developments propose mode-wise normalized driving forces for anisotropic fracture. In particular, \citet{pranavi2024anisotropic} and \citet{pranavi2024unifying} formulate the driving force as an additive combination of fiber and matrix energy contributions, each normalized by the corresponding fracture toughness, with power-law exponents governing mode interaction. While these mathematical objects are general enough for distinguishing between isotropic and anisotropic energy contributions, it turns out that the fiber energy term enters the driving force with a fixed weight regardless of the geometric relationship between the crack orientation and the fiber direction; consequently, the directional sensitivity of fracture resistance to crack--fiber alignment is not captured by power-law exponents alone. In parallel, mixed-mode cohesive phase-field formulations receive growing attention: \citet{wang2024phase} propose a multi-phase-field model with embedded Hashin criteria for fiber-reinforced composites, \citet{efm2025mixed} integrate the Benzeggagh--Kenane (B--K) law directly into a phase-field framework through crack-surface dissipation, \citet{sharma2025combined} combine an explicit cohesive zone model for interface debonding with a phase-field model for matrix cracking, and \citet{bui2025adaptive} develop an adaptive cohesive phase-field model with directional energy decomposition for heterogeneous materials. As a result, a deep understanding of how to combine physically interpretable mode separation with flexible, non-linear interaction laws in a variationally consistent manner is still lacking.

To clarify the physical motivation, it is instructive to consider the two distinct mechanisms through which anisotropy influences fracture. The anisotropic crack density function $\gamma(\nabla d, \mathbf{A})$, governed by the structural tensor $\mathbf{A} = \mathbf{I} + \beta\,\mathbf{a}\otimes\mathbf{a}$, controls the direction-dependent fracture resistance: it penalizes damage gradients along the fiber direction, making it energetically costly to propagate a crack across the fiber and inexpensive to propagate along it. The anisotropic strain energy $\Psi_{\mathrm{ani}}$, which accumulates when the fiber is stretched, governs the direction-dependent fracture driving force: it represents elastic energy stored in the fiber that can be released upon fracture. These two mechanisms are independent and complementary---the crack density controls where the crack prefers to go, while $\Psi_{\mathrm{ani}}$ controls how strongly the crack is driven. A critical question is whether these two mechanisms play the same role across different geometries, or whether their relative importance changes with the geometric setting.

The inclusion of $\Psi_{\mathrm{ani}}$ in the driving force is particularly important for mixed-mode fracture modeling. In isotropic materials, the mode-mixity at a crack tip is determined purely by the geometry---the loading direction and crack orientation---and the isotropic driving force $(\Psi_{\mathrm{v}}^{+}/G_{c,\mathrm{I}})^{p}+(\Psi_{\mathrm{d}}^{+}/G_{c,\mathrm{II}})^{q}$ suffices. In fiber-reinforced materials, however, fibers oriented at an angle to the crack create an asymmetric energy field around the crack tip, generating an intrinsic, {material-induced mode-mixity} even under macroscopically uniaxial loading. Without $\Psi_{\mathrm{ani}}$ in the driving force, this material-induced mode-mixity is lost: the crack density anisotropy alone can deflect the crack path by modifying the fracture resistance, but the driving force magnitude remains isotropic, leading to inaccurate prediction of critical loads and fracture initiation sequences. Including $\Psi_{\mathrm{ani}}$ in the driving force---with its own fracture toughness $G_{c,\mathrm{ani}}$ and interaction exponent $r$---captures this direction-dependent energy release and enables physically consistent mode separation in anisotropic materials. Moreover, in stress-concentration geometries where cracks nucleate from smooth boundaries rather than from pre-existing notches, $\Psi_{\mathrm{ani}}$ modifies the entire elastic strain energy distribution around the stress concentrator, governing not only the critical load but also the fiber-orientation-dependent stiffness and fracture displacement---a geometry-dependent role expansion that is absent in pre-notched geometries.

Although the unified framework combining both mechanisms appears in prior formulations \citep{pranavi2024anisotropic,pranavi2024unifying}, several fundamental questions remain unanswered: (1)~does the anisotropic strain energy produce crack path deflection on its own, or is its role limited to modifying the driving force magnitude? (2)~does the role of each mechanism change between pre-notched and stress-concentration geometries? (3)~when both mechanisms act simultaneously, do they interact linearly or nonlinearly? To the authors' knowledge, no systematic parametric study has addressed these questions. The present work fills this gap through controlled numerical experiments that activate each mechanism independently before combining them.

The present work adopts the AT1 phase-field model \citep{pham2011gradient} with the linear geometric function $\alpha(d)=d$, which possesses a finite elastic threshold ($\alpha'(0)=1>0$). A purely energy-based driving force, however, does not distinguish {how} the crack plane is loaded---whether by opening (mode~I) or by shear (mode~II)---despite the fact that the two modes generally possess different fracture toughnesses ($G_{c,\mathrm{I}}\neq G_{c,\mathrm{II}}$) \citep{benzeggagh1996measurement, rezaei2022anisotropic, pillai2024length}. A physically faithful model must therefore account for the orientation of the crack relative to the material axes, which motivates the orientation-dependent mixed-mode criterion in this work.

The present study employs a single fiber family within the AT1 phase-field framework, proposing a formulation for anisotropic mixed-mode fracture and conducting a systematic parametric study to elucidate the distinct roles of the two anisotropy mechanisms. The central methodological contribution is a normalized mixed-mode crack driving force that separates isotropic and anisotropic contributions: volumetric tensile and deviatoric tensile energies are normalized by the isotropic mode-I and mode-II toughnesses, while the anisotropic fiber energy is normalized by the anisotropic toughness $G_{c,\mathrm{ani}}$. Unlike spectral decomposition methods \citep{miehe2010thermodynamically}, which split the strain energy based on principal strain signs without reference to mode mixity, and unlike stress-based criteria (e.g., Hashin-type) \citep{dean2020anisotropic, mandal2020length}, which define failure envelopes without explicit mode ratios, the present approach provides a physically transparent additive decomposition with tunable power-law exponents ($p,q,r$) governing mode interaction. Furthermore, the history-variable approach is replaced by a bound-constrained minimization, ensuring that damage evolution satisfies thermodynamic irreversibility strictly through the mathematical formulation of the variational inequality.

The major contributions of this work are (i)~a systematic parametric study on single-edge-notched (SEN) and open-hole tension (OHT) specimens that isolates and compares the two anisotropy mechanisms, revealing that the crack density anisotropy controls the crack path to the full fiber angle while the anisotropic strain energy saturates rapidly---indicating that $k_1$ is a material property rather than a tunable parameter---and that the role of $\Psi_{\mathrm{ani}}$ undergoes a geometry-dependent expansion from pre-notched to stress-concentration geometries; (ii)~the discovery of nonlinear synergistic interaction between the two mechanisms, where the combined effect far exceeds the sum of the individual contributions; (iii)~a normalized mixed-mode crack driving force with tunable power-law exponents ($p,q,r$) combining volumetric, deviatoric, and anisotropic contributions each normalized by the corresponding fracture toughness; and (iv)~integration within a finite-strain hyperelastic framework using a generalized neo-Hookean (GNH) deviatoric energy with an AT1 phase-field model, bound-constrained irreversibility, and a $J$-based volumetric split, implemented via a staggered Newton/Trust-Region scheme.

The remainder of this paper is organized as follows. Section~\ref{sec:2} introduces a thermodynamically consistent phase-field model for anisotropic mixed-mode fracture, including kinematics, phase-field regularization, conservation laws, the Clausius--Duhem inequality, the governing equations derived from the Coleman--Noll procedure, the variational formulation with irreversibility, and the anisotropic mixed-mode fracture criterion comprising the anisotropic crack density with structural tensors and the orientation-dependent mixed-mode crack driving force. Section~\ref{sec:3} specifies constitutive choices: the AT1 crack geometric and degradation functions, the volumetric-deviatoric decomposition of the strain energy with anisotropic contribution, and the bound-constrained irreversibility formulation. Section~\ref{sec:4} presents the finite element method, including the weak form, discretization, and the staggered nonlinear solution strategy. Section~\ref{sec:5} demonstrates the framework through numerical examples: a baseline isotropic edge-cracked plate validated against experimental data, a systematic parametric study on single-edge-notched (SEN) specimens that isolates and combines the two anisotropy mechanisms, and an open-hole tension (OHT) study that extends the investigation to stress-concentration geometries where crack nucleation interacts with the anisotropic driving force. Section~\ref{sec:6} concludes with the key findings and remarks on limitations. Appendix~\ref{sec:A} derives the crack evolution equation.

\begin{figure}[!htb]
  \centering
  \includegraphics[width=0.8\textwidth]{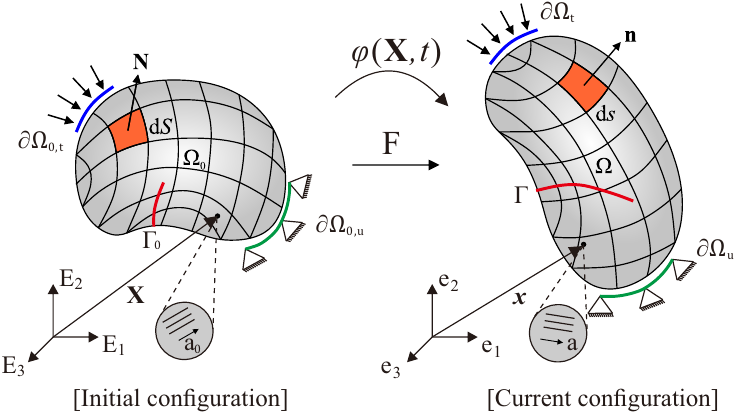}
  \caption{Kinematics of a deformable body with a crack. The deformation mapping $\boldsymbol{\varphi}(\mathbf{X},t)$ maps the reference configuration $\Omega_0$ to the current configuration $\Omega$, with deformation gradient $\mathbf{F}$. The crack surface $\Gamma_0$ in the reference configuration is mapped to $\Gamma$ in the current configuration. Boundaries are decomposed into Dirichlet ($\partial\Omega_{0,u}$) and Neumann ($\partial\Omega_{0,t}$) parts. The fiber direction $\mathbf{a}_0$ in the reference configuration is convected to $\mathbf{a}$ in the current configuration.}
  \label{fig:kinematics}
\end{figure}

\section{A phase-field model for anisotropic mixed-mode fracture}\label{sec:2}
This section introduces a thermodynamically consistent phase-field model for anisotropic mixed-mode fracture. Starting from the conservation laws and the second law of thermodynamics, the governing equations are derived via the Coleman--Noll procedure, ensuring thermodynamic admissibility of the coupled displacement--damage system. The formulation follows a variational structure with anisotropic crack density and a mixed-mode crack driving force. The driving force is composed of volumetric, deviatoric, and per-family anisotropic contributions, each normalized by the corresponding fracture toughness and governed by power-law exponents, aligning with recent anisotropic and mixed-mode developments while preserving variational consistency \citep{wu2020anisotropic, pranavi2024unifying, pillai2024length}.

\subsection{Kinematics}\label{sec:2.1}
Consider a body occupying the reference configuration $\Omega_0 \subset \mathbb{R}^{n_{\mathrm{dim}}}$ with $n_{\mathrm{dim}} \in \{2,3\}$ at time $t_0$. The deformation is described by the mapping $\boldsymbol{\varphi}: \Omega_0 \times [0,T] \to \mathbb{R}^{n_{\mathrm{dim}}}$ such that a material point $\mathbf{X} \in \Omega_0$ is mapped to the spatial position $\mathbf{x}=\boldsymbol{\varphi}(\mathbf{X},t)$ at time $t$. The displacement field is defined as $\mathbf{u}(\mathbf{X},t)=\mathbf{x}(\mathbf{X},t)-\mathbf{X}$. The deformation gradient tensor is
\begin{align}\label{eqn:1}
    \mathbf{F}(\mathbf{X},t) 
    = \nabla_{\mathbf{X}}\boldsymbol{\varphi}(\mathbf{X},t)
    = \mathbf{I} + \nabla_{\mathbf{X}}\mathbf{u}(\mathbf{X},t),
\end{align}
with Jacobian determinant $J=\det\mathbf{F}>0$ for admissible (orientation-preserving) deformations. The right Cauchy--Green deformation tensor is $\mathbf{C} = \mathbf{F}^{\mathrm{T}}\mathbf{F}$. A multiplicative volumetric--isochoric split of the deformation gradient is adopted as $\mathbf{F}=J^{1/3}\bar{\mathbf{F}}$, yielding the isochoric (volume-preserving) right Cauchy--Green tensor $\bar{\mathbf{C}}=J^{-2/3}\mathbf{C}$ with $\det\bar{\mathbf{C}}=1$.

\begin{remark}[Notation]\label{rmk:1}
    Throughout this work, the following conventions are adopted:
    \begin{enumerate}\setlength{\itemsep}{0pt}\setlength{\parsep}{0pt}
        \item Vectors and second-order tensors are denoted by bold symbols (e.g., $\mathbf{u}$, $\mathbf{F}$, $\mathbf{P}$), while scalars are set in italic (e.g., $d$, $J$, $\ell_d$).
        \item The gradient operator $\nabla$ denotes differentiation with respect to material coordinates $\mathbf{X}$ unless otherwise specified. For clarity, the notation $\nabla_{\mathbf{X}}(\cdot)$ is occasionally used.
        \item The dyadic product is denoted by $\otimes$, the dot product by $\cdot$, and the double contraction by $:$, e.g., $\mathbf{A}:\mathbf{B} = A_{ij}B_{ij}$.
        \item For a second-order tensor $\mathbf{A}$, $\mathrm{tr}(\mathbf{A})$ and $\det(\mathbf{A})$ denote its trace and determinant, and $\mathbf{A}^{-\mathrm{T}}$ denotes the inverse transpose.
        \item The Einstein summation convention is adopted: repeated indices imply summation. Unless noted otherwise, Latin indices $i,j,k,\ldots$ take values in $\{1,2,\ldots,n_{\mathrm{dim}}\}$.
        \item The identity tensor is denoted by $\mathbf{I}$. The positive part (Macaulay bracket) is $\langle x \rangle_{+}=\max(x,0)$ for any scalar $x$.
        \item Boundary sets are denoted by $\partial\Omega_{0,u}$ (Dirichlet) and $\partial\Omega_{0,t}$ (Neumann). Prescribed values are indicated by a check accent, e.g., $\check{\mathbf{u}}$ and $\check{d}$.
        \item The superposed dot (e.g., $\dot{\mathbf{A}}$) represents the material time derivative.
        \item Text subscripts use roman font via $\mathrm{}$ command (e.g., $\Psi_{\mathrm{v}}$, $\Psi_{\mathrm{d}}$, $G_{c}$).
    \end{enumerate}
\end{remark}

\begin{figure}[!htb]
  \centering
  \includegraphics[width=0.7\textwidth]{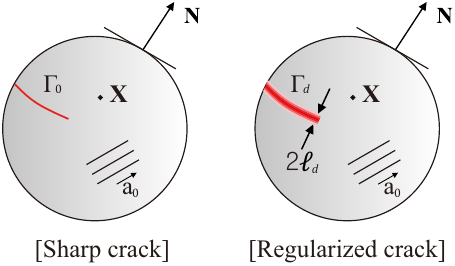}
  \caption{Phase-field regularization of a sharp crack. Left: a body with a sharp crack surface $\Gamma_0$ in the reference configuration. Right: the regularized representation, where the sharp crack is replaced by a diffuse damage band $\Gamma_d$ of width $2\ell_d$ governed by the phase-field variable $d \in [0,1]$. The fiber direction $\mathbf{a}_0$ and the outward normal $\mathbf{N}$ are indicated.}
  \label{fig:regularization}
\end{figure}

\subsection{Phase-field regularization}\label{sec:2.2}
Griffith's energy balance \citep{griffith1921vi} can be written in a variational form \citep{francfort1998revisiting,borden2012phase,miehe2015phase,wu2017unified,kristensen2021assessment,vajari2023investigation,pranavi2024unifying,kim2026unified}:
\begin{align}\label{eqn:2}
    \Pi 
    = \int_{\Omega_0}\Psi_{e}\,\mathrm{d}V 
    + \int_{\Gamma} G_{c}\,\mathrm{d}S,
\end{align}
where $\Psi_{e}$ is the strain energy density and $\Gamma$ is the sharp crack surface with fracture toughness $G_c$. The crack surface energy is regularized by introducing a continuous phase-field variable $d \in [0,1]$, where $d=0$ represents intact material and $d=1$ represents fully broken material. The phase-field regularized functional is
\begin{align}\label{eqn:3}
    \Pi
    \approx \int_{\Omega_0} g(d)\,\Psi_{e}\,\mathrm{d}V
    + \int_{\Omega_0} \Psi_{f}(d, \nabla d)\,\mathrm{d}V,
\end{align}
where $g(d)$ is the degradation function and $\Psi_{f}$ is the fracture energy density, defined as
\begin{align}\label{eqn:psi_f}
    \Psi_{f}(d, \nabla d) = \frac{G_{c}}{c_{w}}\, \gamma_c(d,\nabla d;\mathbf{A},\ell_d).
\end{align}
Here $\gamma_c$ is the anisotropic crack density function, $c_w$ is a normalization constant, $\ell_d$ controls the width of the diffusive crack, and $\mathbf{A}$ is a second-order structural tensor encoding the fiber-reinforced direction; the explicit form of $\gamma_c$ is given in Section~\ref{sec:3.1} and the construction of $\mathbf{A}$ is detailed in Section~\ref{sec:2.7}.

\subsection{Balance principles}\label{sec:2.3}
The governing equations consist of two balance principles: the macroscopic mechanical equilibrium and the microscopic crack evolution.

The macroscopic balance is governed by the balance of linear momentum in the reference configuration:
\begin{alignat}{2} 
    \nabla\cdot\mathbf{P} + \mathbf{B} &= 0 \quad &&\text{in} \quad \Omega_0, \label{eqn:4} \\
    \mathbf{u} &= \check{\mathbf{u}} \quad &&\text{on} \quad \partial\Omega_{0,u}, \label{eqn:5} \\
    \mathbf{P}\cdot\mathbf{N} &= \mathbf{T} \quad &&\text{on} \quad \partial\Omega_{0,t}, \label{eqn:6}
\end{alignat}
where $\mathbf{P}$ is the first Piola--Kirchhoff stress tensor, $\mathbf{B}$ is the body force per unit reference volume, $\check{\mathbf{u}}$ is the prescribed displacement on the Dirichlet boundary $\partial\Omega_{0,u}$, $\mathbf{T}$ is the prescribed traction on the Neumann boundary $\partial\Omega_{0,t}$, and $\mathbf{N}$ is the outward unit normal.

The microscopic balance governing crack evolution involves the scalar microstress $\omega$ and vector microstress $\boldsymbol{\xi}$ \citep{wu2017unified,mao2018theory}:
\begin{alignat}{2} 
    \nabla\cdot\boldsymbol{\xi} - \omega &= 0 \quad &&\text{in} \quad \Omega_0, \label{eqn:7} \\ 
    d &= \check{d} \quad &&\text{on} \quad \Gamma_0, \label{eqn:8}
\end{alignat}
where $\omega$ is the scalar microstress conjugate to the damage variable $d$, $\boldsymbol{\xi}$ is the vector microstress conjugate to the damage gradient $\nabla d$, and $\check{d}$ is the prescribed damage on the boundary $\Gamma_0$.

\subsection{Thermodynamics and constitutive relations}\label{sec:2.4}
The energy conservation for the coupled displacement--damage problem combines the macroscopic and microscopic power expenditures, with the local rate of internal energy given by $\dot{\mathnormal{W}} = \mathbf{P}:\dot{\mathbf{F}} + \omega\,\dot{d} + \boldsymbol{\xi}\cdot\nabla\dot{d}$. Since the failure process involves energy dissipation within the fracture process zone, this dissipative system must be complemented by the second law of thermodynamics to guarantee thermodynamic admissibility \citep{coleman1963thermodynamics, boo2025multiphysics}.
For isothermal quasi-static processes, the Clausius--Duhem inequality requires that the rate of dissipation be non-negative. Introducing the Helmholtz free energy density $\Psi = \mathnormal{W} - T\eta$, where $T$ is the absolute temperature and $\eta$ is the entropy per unit reference volume, and substituting the energy balance, the local form of the Clausius--Duhem inequality reads
\begin{align}
    \mathbf{P}:\dot{\mathbf{F}} + \omega\,\dot{d} + \boldsymbol{\xi}\cdot\nabla\dot{d} - \dot{\Psi} \geq 0. \label{eqn:CD_local}
\end{align}

In the coupled displacement--damage system, the Helmholtz free energy density can be decomposed into the elastic strain energy density $\Psi_{e}$ and the fracture energy density $\Psi_{f}$, as
\begin{align}
    \Psi = \Psi_{e}(\mathbf{F}, d) + \Psi_{f}(d, \nabla d), \label{eqn:9a}
\end{align}
where $\Psi_{e}(\mathbf{F}, d) = g(d)\,\Psi_{\mathrm{und}}(\mathbf{F})$ with $g(d)$ being the degradation function and $\Psi_{\mathrm{und}}(\mathbf{F})$ the undamaged strain energy density, and $\Psi_{f}$ is the fracture energy density defined in \eqref{eqn:psi_f}.

Accordingly, the rate of change of the Helmholtz free energy density, $\dot{\Psi}$, can be given by
\begin{align}
    \dot{\Psi} = \frac{\partial \Psi_{e}}{\partial \mathbf{F}}:\dot{\mathbf{F}}
    + \left(\frac{\partial \Psi_{e}}{\partial d} + \frac{\partial \Psi_{f}}{\partial d}\right)\dot{d}
    + \frac{\partial \Psi_{f}}{\partial \nabla d}\cdot\nabla\dot{d}. \label{eqn:H_rate}
\end{align}
Substituting \eqref{eqn:H_rate} into \eqref{eqn:CD_local}, the local Clausius--Duhem inequality can be rewritten as
\begin{align}
    \left(\mathbf{P} - \frac{\partial \Psi_{e}}{\partial \mathbf{F}}\right):\dot{\mathbf{F}}
    + \left(\omega - \frac{\partial \Psi_{e}}{\partial d} - \frac{\partial \Psi_{f}}{\partial d}\right)\dot{d}
    + \left(\boldsymbol{\xi} - \frac{\partial \Psi_{f}}{\partial \nabla d}\right)\cdot\nabla\dot{d}
    \geq 0. \label{eqn:CD_expanded}
\end{align}
The inequality in \eqref{eqn:CD_expanded} must hold for arbitrary thermodynamic processes, which demands that the dissipative terms be non-negative and the coefficients of the non-dissipative terms vanish. In this study, it is assumed that the entropy is produced solely by the creation of new crack surfaces. Therefore, \eqref{eqn:CD_expanded} can be divided into the non-dissipative terms, which yield the constitutive relations:
\begin{align}\label{eqn:micro_constitutive}
    \mathbf{P} = \frac{\partial \Psi_{e}}{\partial \mathbf{F}}, \quad
    \omega = \dfrac{\partial \Psi_{e}}{\partial d} + \dfrac{\partial \Psi_{f}}{\partial d}, \quad \text{and} \quad
    \boldsymbol{\xi} = \dfrac{\partial \Psi_{f}}{\partial \nabla d}.
\end{align}
Since the fracture energy density $\Psi_{f}$ in \eqref{eqn:psi_f} already accounts for the total energy required to create crack surfaces through the fracture toughness $G_c$, the microstresses $\omega$ and $\boldsymbol{\xi}$ are taken to be purely energetic (non-dissipative). The thermodynamic admissibility of the fracture process is then enforced by the damage irreversibility constraint $\dot{d} \geq 0$ together with the variational structure of the governing equations.

\subsection{Strain energy decomposition}
For the displacement problem, the elastic strain energy density $\Psi_{e}$ in \eqref{eqn:9a} can be decomposed into two parts, based on whether that part of the energy density can drive the evolution of damage, as
\begin{align}
    \Psi_{e}(\mathbf{F}, d) = g(d)\,\Psi^{+}(\mathbf{F}) + \Psi^{-}(\mathbf{F}), \label{eqn:psi_e_split}
\end{align}
where $\Psi^{+}$ is the elastic strain energy density part that drives the evolution of damage and $\Psi^{-}$ is the remaining part. The tension--compression decomposition $\Psi_{\mathrm{und}} = \Psi^{+} + \Psi^{-}$ is detailed in Section~\ref{sec:3.2}.

Accordingly, substituting \eqref{eqn:psi_e_split} into \eqref{eqn:micro_constitutive}, the constitutive relationship of the displacement problem can be obtained as
\begin{align}
    \mathbf{P} = \frac{\partial \Psi_{e}}{\partial \mathbf{F}} = g(d)\,\frac{\partial \Psi^{+}}{\partial \mathbf{F}} + \frac{\partial \Psi^{-}}{\partial \mathbf{F}}. \label{eqn:P_split}
\end{align}

For the damage problem, substituting the energy densities \eqref{eqn:9a} and \eqref{eqn:psi_f} into the constitutive relations \eqref{eqn:micro_constitutive} gives the explicit forms:
\begin{align}
    \omega &= g'(d)\,\Psi^{+} + \dfrac{G_{c}}{c_{w}}\dfrac{\alpha'(d)}{2\ell_{d}}, \label{eqn:micro_constitutive_explicit_a} \\
    \boldsymbol{\xi} &= \dfrac{G_{c}}{c_{w}}\,\ell_{d}\,\mathbf{A}\,\nabla d. \label{eqn:micro_constitutive_explicit_b}
\end{align}

\subsection{Phase-field evolution equation}
Substituting \eqref{eqn:micro_constitutive_explicit_a} and \eqref{eqn:micro_constitutive_explicit_b} into the micro-force balance \eqref{eqn:7} yields the phase-field evolution equation in terms of the crack driving force $\mathcal{H}$, whose explicit form is given in Section~\ref{sec:2.8} (Eq.~\eqref{eqn:anis_H}):
\begin{align}
    g'(d)\,\mathcal{H} + \frac{1}{c_{w}}\left[\frac{\alpha'(d)}{2\ell_{d}} - \ell_{d}\,\nabla\cdot\left(\mathbf{A}\,\nabla d\right)\right] = 0. \label{eqn:pf_evolution}
\end{align}

The irreversibility of damage requires $\dot{d} \geq 0$, so that the material cannot heal. This condition can be enforced by introducing a history variable that tracks the maximum crack driving force $\mathcal{H}$ over time \citep{miehe2010thermodynamically}, or by formulating the problem as a variational inequality with bound constraints on $d$ \citep{bourdin2000numerical}. The present work adopts the latter approach, which preserves the variational structure and avoids the need for auxiliary history fields. Crack irreversibility is enforced by the bound constraint
\begin{align}
    d_{n-1} \le d \le 1 \quad \text{in} \quad \Omega_0, \label{eqn:anis_bounds}
\end{align}
so that damage can only increase between load steps. A detailed derivation of the phase-field evolution equation~\eqref{eqn:pf_evolution} and the irreversibility conditions~\eqref{eqn:anis_bounds} is provided in Appendix~\ref{sec:A}.

\subsection{Anisotropic structural tensor}\label{sec:2.7}
The structural tensor $\mathbf{A}$ (cf.\ Eq.~\eqref{eqn:psi_f}) for a single fiber family with unit direction vector $\mathbf{a}$ is defined as
\begin{align}
    \mathbf{A} = \mathbf{I} + \beta\,\mathbf{a}\otimes\mathbf{a}, \label{eqn:anis_A}
\end{align}
where $\beta \geq 0$ controls the anisotropy strength. When $\beta = 0$, the isotropic case $\mathbf{A}=\mathbf{I}$ is recovered. The concrete parametrization of $\mathbf{a}$ for the two-dimensional setting is given in Section~\ref{sec:3.3}.

\subsection{Mixed-mode crack driving force}\label{sec:2.8}

The normalized mixed-mode crack driving force is defined as
\begin{align}
    \mathcal{H} &=
    \left(\frac{\Psi_{\mathrm{v}}^{+}}{G_{c,\mathrm{I}}}\right)^{p}
    + \left(\frac{\Psi_{\mathrm{d}}^{+}}{G_{c,\mathrm{II}}}\right)^{q}
    + \left(\frac{\Psi_{\mathrm{ani}}^{+}}{G_{c,\mathrm{ani}}}\right)^{r}, \label{eqn:anis_H}
\end{align}
where $\Psi_{\mathrm{ani}}^{+}$ is the tensile anisotropic strain energy, $G_{c,\mathrm{I}}$ and $G_{c,\mathrm{II}}$ are the mode-I and mode-II fracture toughnesses of the matrix, $G_{c,\mathrm{ani}}$ is the anisotropic fracture toughness, and $p$, $q$, $r$ are power-law exponents governing nonlinear mode interaction ($p=q=r=1$ recovers linear superposition). The first two terms represent the isotropic contributions from volumetric tension and deviatoric tensile stretching, each normalized by the corresponding toughness. The third term represents the anisotropic contribution normalized by $G_{c,\mathrm{ani}}$.

\section{Specific material models}\label{sec:3}
This section specifies the constitutive choices that specialize the general framework. The strain energy consists of an isochoric generalized neo-Hookean (GNH) deviatoric base with an invariant-based tension--compression split, a a $J$-based volumetric energy and an anisotropic fiber energy aligned with the material axis. The mixed-mode crack driving force is assembled from mode-wise normalized energies and tunable interaction exponents, allowing a calibrated transition between linear and nonlinear coupling \citep{pranavi2024anisotropic, pranavi2024unifying, pillai2024length}.
Specific selections are made for the strain energy density, crack geometrical functions, and strain energy decomposition.

\subsection{Crack geometric functions and degradation function}\label{sec:3.1}

The present work adopts the AT1 phase-field model \citep{pham2011gradient}, which employs a linear crack geometric function together with a standard quadratic degradation function.

The anisotropic crack density function appearing in \eqref{eqn:psi_f} is defined as
\begin{align}
    \gamma_c(d,\nabla d;\mathbf{A},\ell_d) = \frac{\alpha(d)}{2\ell_{d}} + \frac{\ell_{d}}{2}\,\nabla d \cdot \mathbf{A}\,\nabla d, \label{eqn:anis_gamma}
\end{align}
where the first term accounts for the local damage through the geometric function $\alpha(d)$ and the second term penalizes spatial gradients of $d$ in an orientation-dependent fashion via the structural tensor $\mathbf{A}$. In this work, the AT1 linear geometric function is adopted:
\begin{align}
    \alpha(d) = d, \label{eqn:alpha_czm}
\end{align}
which satisfies $\alpha(0)=0$, $\alpha(1)=1$, and crucially $\alpha'(0)=1>0$, ensuring a finite elastic threshold below which no damage evolves. The normalization constant becomes
\begin{align}
    c_{w} = 4\int_{0}^{1}\sqrt{\alpha(z)}\,\mathrm{d}z = 4\int_{0}^{1}\sqrt{z}\,\mathrm{d}z = \frac{8}{3}. \label{eqn:cw_czm}
\end{align}

The standard quadratic degradation function is adopted:
\begin{align}
    g(d) = (1-d)^{2}, \label{eqn:g_at1}
\end{align}
which satisfies $g(0)=1$, $g(1)=0$, and $g'(d)\le 0$ for $d\in[0,1]$. It is worth noting that the AT1 model differs from the AT2 model in the choice of the geometric function: the AT2 model adopts $\alpha(d)=d^2$, for which $\alpha'(0)=0$, resulting in $c_w=2$ and no elastic threshold---damage initiates immediately upon loading. In contrast, the AT1 linear function $\alpha(d)=d$ with $\alpha'(0)=1>0$ ensures a finite elastic threshold, since damage does not evolve until the driving force exceeds a critical value. Both models share the same quadratic degradation function $g(d)=(1-d)^2$.

\begin{remark}[The phase-field length scale $\ell_d$]\label{rmk:lengthscale}
The regularization length $\ell_d$ in the crack density functional~\eqref{eqn:anis_gamma} serves four interrelated purposes:
\begin{enumerate}
    \item \textbf{Diffuse crack width.} $\ell_d$ controls the half-width of the damage band. For the AT1 model the damage profile has compact support with a finite damage band width.
    \item \textbf{Strength.} For the AT1 model, the critical stress of the homogeneous 1D bar under uniaxial tension with Young's modulus $E$ is
    \begin{align*}
        \sigma_c = \sqrt{\frac{3\,G_c\,E}{8\,\ell_d}}\,.
    \end{align*}
    The AT1 model possesses a finite elastic threshold ($\alpha'(0)=1>0$): the material responds purely elastically until the stress reaches $\sigma_c$, after which damage initiates.
    \item \textbf{Mesh requirement.} Resolving the damage profile requires a mesh size $h$ satisfying $h \le \ell_d/2$ within the damage zone \citep{borden2012phase}. If this condition is violated, the effective fracture toughness is over-estimated and damage initiation is delayed.
    \item \textbf{$\Gamma$-convergence.} As $\ell_d \to 0$, the regularized functional $\Gamma$-converges to the Griffith surface energy, recovering the sharp-crack theory \citep{bourdin2000numerical}.
\end{enumerate}
In practice, $\ell_d$ is set as a small multiple of the minimum element size, $\ell_d = m \cdot h_{\min}$, where the multiplier $m \ge 2$ ensures the mesh resolution criterion $h/\ell_d \le 1/2$ is satisfied.
\end{remark}

\subsection{Strain energy density}\label{sec:3.2}
The present formulation employs an isochoric generalized neo-Hookean (GNH) deviatoric energy \citep{horgan2004constitutive,lu2021mixed} with an invariant-based tension--compression split, combined with a $J$-based volumetric split and an additional anisotropic contribution aligned with the material axis. A prime denotes differentiation with respect to the phase-field variable $d$, i.e., $g'(d)=\mathrm{d}g/\mathrm{d}d$ and $\alpha'(d)=\mathrm{d}\alpha/\mathrm{d}d$.

The undamaged strain energy density is decomposed into isotropic and anisotropic contributions:
\begin{align}
    \Psi(J,\bar{\mathbf{C}};\mathbf{a}) = \Psi_{\mathrm{iso}} + \Psi_{\mathrm{ani}}, \label{eqn:psi_total}
\end{align}
where the isotropic part consists of volumetric and deviatoric contributions:
\begin{align}
    \Psi_{\mathrm{iso}} =
    \underbrace{\frac{\kappa}{2}\left(J-1\right)^{2}}_{\Psi_{\mathrm{v}}}
    + \underbrace{\frac{\mu}{2b}\left\{\left[1+\frac{b}{n}\left(\bar{I}_1-3\right)\right]^{n}-1\right\}}_{\Psi_{\mathrm{d}}}, \label{eqn:psi_iso}
\end{align}
$\kappa$ and $\mu$ are the bulk modulus and shear modulus of the matrix material, respectively, $\bar{I}_1 = \mathrm{tr}\,\bar{\mathbf{C}} = \sum_{j=1}^{3}\bar{\lambda}_j^2$ is the first invariant of the isochoric right Cauchy--Green tensor $\bar{\mathbf{C}}=J^{-2/3}\mathbf{C}$, $b>0$ is a dimensionless hardening parameter that controls the onset of strain stiffening, and $n\ge 1$ is the exponent of the generalized neo-Hookean (GNH) model \citep{horgan2004constitutive,lu2021mixed}: $n=1$ recovers the standard neo-Hookean model $\Psi_{\mathrm{d}}=(\mu/2)(\bar{I}_1-3)$ (the $b$-dependence cancels), while $n\ge 2$ introduces strain stiffening at large deformation with the same initial shear modulus $G_0=\mu$. The anisotropic energy is
\begin{align}
    \Psi_{\mathrm{ani}} = \frac{k_{1}}{k_{2}}\left[\exp\!\left(k_{2}\left\langle\bar{\mathbf{C}}:\mathbf{a}\otimes\mathbf{a}-1\right\rangle\right)-1\right] - k_{1}\left\langle\bar{\mathbf{C}}:\mathbf{a}\otimes\mathbf{a}-1\right\rangle. \label{eqn:psi_ani_i}
\end{align}
$k_{1}$ and $k_{2}$ represent the stiffness and exponent of the fiber reinforcement, and $\langle w \rangle = \max(w,0)$ denotes the Macaulay bracket.

\subsection{Spectral decomposition}
In the general three-dimensional setting, the eigenvalues $\lambda_1^2,\lambda_2^2,\lambda_3^2$ of $\mathbf{C}=\mathbf{F}^{\mathrm{T}}\mathbf{F}$ are obtained from the characteristic equation
\begin{align}
    \det\!\left(\mathbf{C} - \lambda_i^2\,\mathbf{I}\right) = 0, \quad i=1,2,3, \label{eqn:eigenvalues_C_3D}
\end{align}
which yields a cubic polynomial in $\lambda_i^2$ whose coefficients are the principal invariants $I_1=\mathrm{tr}\,\mathbf{C}$, $I_2=\tfrac{1}{2}[(\mathrm{tr}\,\mathbf{C})^2-\mathrm{tr}(\mathbf{C}^2)]$, and $I_3=\det\mathbf{C}$.
The three real roots are given explicitly by the trigonometric solution \citep{smith1961eigenvalues,basar1998finite}
\begin{align}
    \lambda_i^2 = \frac{1}{3}\!\left(I_1(\mathbf{C}) + 2\sqrt{I_1^2(\mathbf{C}) - 3I_2(\mathbf{C})}\;\cos\!\left(\frac{\vartheta + 2\pi\,(i-1)}{3}\right)\!\right), \quad i=1,2,3, \label{eqn:eigenvalues_trig}
\end{align}
with the Lode angle
\begin{align}
    \vartheta = \arccos\!\left(\frac{2I_1^3(\mathbf{C}) - 9I_1(\mathbf{C})I_2(\mathbf{C}) + 27I_3(\mathbf{C})}{2\left(I_1^2(\mathbf{C}) - 3I_2(\mathbf{C})\right)^{3/2}}\right). \label{eqn:lode_angle}
\end{align}
The isochoric eigenvalues follow as $\bar{\lambda}_i^2 = J^{-2/3}\lambda_i^2$.
This closed-form expression avoids iterative eigensolvers and is amenable to automatic differentiation within finite element frameworks.
The two-dimensional specialization is given in Section~\ref{sec:3.3}.

\subsection{Strain energy decomposition}
The tension--compression split is applied to the volumetric energy, the isochoric deviatoric energy, and the anisotropic energy. The deviatoric energy $\Psi_{\mathrm{d}}=(\mu/2b)\{[1+(b/n)(\bar{I}_1-3)]^n-1\}$ couples all principal stretches through $\bar{I}_1$, and an invariant-based tension--compression split on $\bar{I}_1$ is adopted.

The positive and negative parts of $\Psi_{\mathrm{v}}$ are
\begin{align}
    \Psi_{\mathrm{v}}^{+} = \frac{\kappa}{2}\left[J-1\right]^{2}\Big|_{J>1}, \qquad
    \Psi_{\mathrm{v}}^{-} = \frac{\kappa}{2}\left[J-1\right]^{2}\Big|_{J<1}, \label{eqn:psi_v_split}
\end{align}
so that volumetric expansion ($J > 1$) drives damage through $\Psi_{\mathrm{v}}^{+}$, while volumetric compression ($J \le 1$) is preserved in $\Psi_{\mathrm{v}}^{-}$. The deviatoric energy is split based on the sign of $\bar{I}_1-3$:
\begin{align}
    \Psi_{\mathrm{d}}^{+} = \mathrm{H}(\bar{I}_1 - 3)\,\Psi_{\mathrm{d}}, \qquad
    \Psi_{\mathrm{d}}^{-} = \mathrm{H}(3 - \bar{I}_1)\,\Psi_{\mathrm{d}}, \label{eqn:psi_d_split}
\end{align}
where $\mathrm{H}(\cdot)$ is the Heaviside step function. Since $\Psi_{\mathrm{d}}(\bar{I}_1{=}3)=0$, the split is continuous at the undeformed state.

The anisotropic energy is activated only when the isochoric fiber stretch exceeds unity:
\begin{align}
    \Psi_{\mathrm{ani}}^{+} =
    \begin{cases}
        \Psi_{\mathrm{ani}}, & \bar{\mathbf{C}}:\mathbf{a}\otimes\mathbf{a} > 1,\\
        0, & \text{otherwise}.
    \end{cases} \label{eqn:psi_ani_split}
\end{align}

The tension--compression split is assembled as
\begin{align}
    \Psi^{+} &= \Psi_{\mathrm{v}}^{+} + \Psi_{\mathrm{d}}^{+} + \Psi_{\mathrm{ani}}^{+}, \label{eqn:psi_plus_total}\\
    \Psi^{-} &= \Psi_{\mathrm{v}}^{-} + \Psi_{\mathrm{d}}^{-}. \label{eqn:psi_minus_total}
\end{align}
Volumetric expansion ($J>1$) and net isochoric extension ($\bar{I}_1 > 3$) drive damage through $\Psi^{+}$, while volumetric compression ($J\le 1$) and net isochoric compression ($\bar{I}_1 \le 3$) are preserved in $\Psi^{-}$ to prevent material interpenetration. The phase-field regularized strain energy density is therefore
\begin{align}
    \Psi(\mathbf{F},d) = g(d)\,\Psi^{+} + \Psi^{-}. \label{eqn:17}
\end{align}
The first Piola--Kirchhoff stress tensor follows as $\mathbf{P} = \partial\Psi/\partial\mathbf{F}$.

\begin{remark}[Mode-wise decomposition for mixed-mode criterion]\label{rmk:mode_decomp}
The spectral split combined with the volumetric energy provides a natural basis for the mixed-mode crack driving force $\mathcal{H}$ (Section~\ref{sec:2.8}). The volumetric tensile energy $\Psi_{\mathrm{v}}^{+}$ serves as a proxy for mode~I (opening) and is normalized by $G_{c,\mathrm{I}}$, while the tensile isochoric deviatoric energy $\Psi_{\mathrm{d}}^{+}$ serves as a proxy for mode~II (shear) and is normalized by $G_{c,\mathrm{II}}$. The anisotropic term $\Psi_{\mathrm{ani}}^{+}$ captures the fiber contribution and is normalized by $G_{c,\mathrm{ani}}$.
\end{remark}

\subsection{Specialization to 2D plane stress problems}\label{sec:3.3}

For two-dimensional problems, the deformation mapping $\boldsymbol{\varphi}$ is symmetric about the mid-plane $X_3=0$, and the out-of-plane shear components vanish: $F_{13}=F_{23}=F_{31}=F_{32}=0$ on the mid-plane. The deformation gradient therefore takes the form
\begin{align}\label{eqn:F_planestress}
    \mathbf{F} = \begin{pmatrix} F_{11} & F_{12} & 0 \\ F_{21} & F_{22} & 0 \\ 0 & 0 & F_{33} \end{pmatrix}, \qquad
    \mathbf{F}_{\mathrm{2D}} = \begin{pmatrix} F_{11} & F_{12} \\ F_{21} & F_{22} \end{pmatrix},
\end{align}
where $F_{33}=\lambda_3$ is the out-of-plane stretch. Under the plane stress assumption adopted in this work, $\lambda_3$ is an additional unknown determined by the stress-free condition $P_{33}=0$ in the thickness direction. This nonlinear equation is solved numerically within the displacement solution step (see Algorithm~\ref{alg:euclid}).

The condition $P_{33}=0$ applied to the phase-field regularized energy $\Psi(\mathbf{F},d)=g(d)\,\Psi^{+}+\Psi^{-}$ reads
\begin{align}\label{eqn:P33}
    P_{33} = \frac{\partial\Psi}{\partial\lambda_3}
    = g(d)\,\frac{\partial\Psi^{+}}{\partial\lambda_3}
    + \frac{\partial\Psi^{-}}{\partial\lambda_3} = 0.
\end{align}
For the present GNH model, $\Psi$ depends on $\lambda_3$ through the volumetric energy (via $J = \det\mathbf{F}_{\mathrm{2D}}\cdot\lambda_3$), the isochoric deviatoric energy (via $\bar{I}_1 = \sum_i \bar{\lambda}_i^2$, where $\bar{\lambda}_i = J^{-1/3}\lambda_i$), and the anisotropic energy (via $\bar{\mathbf{C}}:\mathbf{a}\otimes\mathbf{a}$). This yields a nonlinear equation in $\lambda_3$ that is solved numerically at each Newton--Raphson iteration of the displacement subproblem.

The block structure \eqref{eqn:F_planestress} implies that $\lambda_3^2$ is directly one eigenvalue of $\mathbf{C}$. The remaining two eigenvalues are those of the in-plane block $\mathbf{C}_{\mathrm{2D}} = \mathbf{F}_{\mathrm{2D}}^{\mathrm{T}}\mathbf{F}_{\mathrm{2D}}$, where $\mathbf{F}_{\mathrm{2D}}$ is the $2\times 2$ upper-left submatrix of $\mathbf{F}$. For a $2\times 2$ symmetric positive-definite matrix, the eigenvalues are computed analytically as
\begin{align}
    \lambda_{1,2}^2 = \frac{\mathrm{tr}\,\mathbf{C}_{\mathrm{2D}}}{2} \pm \sqrt{\left(\frac{\mathrm{tr}\,\mathbf{C}_{\mathrm{2D}}}{2}\right)^2 - \det\mathbf{C}_{\mathrm{2D}}}. \label{eqn:eigenvalues_C}
\end{align}
The three eigenvalues $\lambda_1^2, \lambda_2^2, \lambda_3^2$ yield the isochoric principal stretches $\bar{\lambda}_i = J^{-1/3}\lambda_i$ and the first invariant $\bar{I}_1 = \sum_i \bar{\lambda}_i^2$. This closed-form evaluation avoids iterative eigensolvers and ensures exact differentiability within the automatic differentiation framework.

In the two-dimensional setting, the fiber direction vector lies in the plane and is parametrized by
\begin{align}
    \mathbf{a} = [\cos\theta_f,\; \sin\theta_f,\; 0]^{\mathrm{T}}, \label{eqn:fiber_direction_2D}
\end{align}
where $\theta_f$ is the fiber orientation angle measured from the reference axis.

\section{Finite element method}\label{sec:4}
This section presents the finite element formulation and nonlinear solution strategy for the phase-field fracture model developed in Sections~\ref{sec:2} and \ref{sec:3}. The implementation uses FEniCS \citep{LoggMardalEtAl2012a,AlnaesBlechta2015a,kim2024finite} with a staggered scheme (alternate minimization) in which the displacement and phase-field subproblems are solved sequentially. This approach is widely adopted in phase-field fracture due to its robustness and algorithmic simplicity, and is consistent with variationally derived weak forms for anisotropic damage and mixed-mode crack driving forces \citep{wu2020anisotropic, kristensen2021assessment}. The section begins with the variational formulation and weak form of the governing equations, followed by the finite element discretization and the nonlinear solution strategy. The formulation below is presented for the two-dimensional plane stress setting of Section~\ref{sec:3.3}. The primary unknowns are the displacement $\mathbf{u}$ and the damage field $d$.

\subsection{Variational formulation}\label{sec:4.1}

The elastic equilibrium~\eqref{eqn:P_split}, the phase-field evolution equation~\eqref{eqn:pf_evolution}, and the irreversibility constraint~\eqref{eqn:anis_bounds} can equivalently be obtained from a constrained minimization of the total potential energy functional
\begin{align}
    \Pi_{0}(\mathbf{u},d)
    = \int_{\Omega_0} \left[g(d)\,G_{c}\mathcal{H} + \Psi_{\mathrm{v}}^{-}(\mathbf{F})\right]\mathrm{d}\Omega_0
    + \int_{\Omega_0} \frac{G_{c}}{c_{w}}\,\gamma_c(d,\nabla d;\mathbf{A},\ell_d)\,\mathrm{d}\Omega_0
    - \mathcal{W}_{\mathrm{ext}}(\mathbf{u}), \label{eqn:anis_Pi0}
\end{align}
where the first integral contains the degraded elastic driving energy and the undegraded compressive part, the second integral is the crack-surface energy scaled by the fracture toughness $G_{c}$, and $\mathcal{W}_{\mathrm{ext}}$ is the external work of applied loads. The equilibrium state is found by seeking the stationary point of $\Pi_0$ with respect to $\mathbf{u}$ and the minimum with respect to $d$ subject to \eqref{eqn:anis_bounds}:
\begin{align}
    \delta_{\mathbf{u}}\Pi_0 = 0, \qquad \delta_{d}\Pi_0 \geq 0 \quad \text{for all admissible } \delta d \geq 0. \label{eqn:variational_ineq}
\end{align}
The first condition recovers the elastic equilibrium~\eqref{eqn:P_split}, while the second, together with the bound constraint~\eqref{eqn:anis_bounds}, yields the phase-field evolution equation~\eqref{eqn:pf_evolution} as the corresponding Karush--Kuhn--Tucker condition (Appendix~\ref{sec:A}).

\subsection{Weak form}\label{sec:4.2}

The weak form is obtained by taking the first variation of the total potential energy in \eqref{eqn:anis_Pi0} with respect to the displacement and the phase-field, using admissible test functions that satisfy the necessary integrability conditions \citep{hughes2012finite}. Multiplying the balance equations with the test functions $\delta\mathbf{u}$ and $\delta d$, and integrating over the domain yields
\begin{align}
    \int_{\Omega_0} \mathbf{P}:\nabla\delta\mathbf{u} \, \mathrm{d}V = \int_{\Omega_0} \mathbf{B}\cdot\delta\mathbf{u}\,\mathrm{d}V + \int_{\partial\Omega_{0,t}} \mathbf{T}\cdot\delta\mathbf{u}\,\mathrm{d}A, \label{eqn:20} \\
    \int_{\Omega_0} \left[g'(d)\,\mathcal{H}\,\delta d
    + \frac{1}{c_{w}} \left(\frac{1}{2\ell_{d}}\,\delta d + \ell_{d}\,\nabla\delta d \cdot \mathbf{A}\,\nabla d\right) \right] \, \mathrm{d}V \geq 0. \label{eqn:21}
\end{align}
Here the normalized crack driving force $\mathcal{H}$ is evaluated according to the mixed-mode decomposition described in Section~\ref{sec:2.8}. The inequality in \eqref{eqn:21} reflects the irreversibility constraint $\dot{d}\geq 0$, whose enforcement via bound constraints is discussed in Section~\ref{sec:4.4}.

The statement of the weak form is to find the trial functions $\{\mathbf{u},d\} \in \mathcal{S}_{u}\times\mathcal{S}_{d}$ such that Equations \eqref{eqn:20} and \eqref{eqn:21} are satisfied for any permissible test functions $\{\delta\mathbf{u},\delta d\} \in \mathcal{V}_{u}\times\mathcal{V}_{d}$. The sets of admissible trial functions are
\begin{align}
    \mathcal{S}_{u} &= \{\mathbf{u}\,\vert\,\mathbf{u}\in[H^{1}(\Omega_0)]^{n_{\mathrm{dim}}}, \; \mathbf{u}=\check{\mathbf{u}} \;\text{on}\;\partial\Omega_{0,u} \}, \label{eqn:22} \\
    \mathcal{S}_{d} &= \{d\,\vert\,d\in H^{1}(\Omega_0), \; d=\check{d}\;\text{on}\;\Gamma_0, \; d \in [0,1] \}, \label{eqn:23}
\end{align}
where $\check{\mathbf{u}}$ and $\check{d}$ are the prescribed values on Dirichlet boundary conditions, and $H^1$ denotes the Sobolev space of degree one. Similarly, the sets of all admissible test functions are
\begin{align}
    \mathcal{V}_{u} &= \{\delta\mathbf{u}\,\vert\,\delta\mathbf{u}\in[H^{1}(\Omega_0)]^{n_{\mathrm{dim}}}, \; \delta\mathbf{u}=\mathbf{0}\;\text{on}\;\partial\Omega_{0,u} \}, \label{eqn:24} \\
    \mathcal{V}_{d} &= \{\delta d\,\vert\,\delta d\in H^{1}(\Omega_0), \; \delta d=0\;\text{on}\;\Gamma_0, \; \delta d \geq 0 \}. \label{eqn:25}
\end{align}

\subsection{Spatial discretization}\label{sec:4.3}
A standard Galerkin discretization is used with continuous, piecewise-linear elements for the displacement and damage fields. The nodal fields are interpolated as
\begin{align}
    \mathbf{u}^h(\mathbf{X},t)
    &= \sum_{a=1}^{n_u}\mathbf{N}^{u}_{a}(\mathbf{X})\mathbf{u}_{a}(t)
    ,\quad
    d^h(\mathbf{X},t)
    = \sum_{a=1}^{n_d}N^{d}_{a}(\mathbf{X})d_{a}(t), \label{eqn:26}
\end{align}
where $N_a$ denotes the shape functions and subscript $a$ denotes node indices. The corresponding gradients are
\begin{align}
    \nabla\mathbf{u}^h(\mathbf{X},t)
    &= \sum_{a=1}^{n_u}\mathbf{B}^{u}_{a}(\mathbf{X})\mathbf{u}_{a}(t)
    ,\quad
    \nabla d^h(\mathbf{X},t)
    = \sum_{a=1}^{n_d}\mathbf{B}^{d}_{a}(\mathbf{X})d_{a}(t), \label{eqn:27}
\end{align}
where $\mathbf{B}_a = \nabla N_a$ are the gradients of the shape functions.

\subsection{Temporal discretization}\label{sec:4.4}

The total potential energy is non-convex with respect to $(\mathbf{u}, d)$ jointly; however, each subproblem is convex when the other field is held fixed \citep{miehe2010thermodynamically}. A staggered scheme (alternate minimization) is therefore adopted. At each load step $n \to n+1$, the displacement field $\mathbf{u}_{n+1}$ is first obtained by solving the elastic equilibrium \eqref{eqn:20} with the damage field frozen at $d_n$:
\begin{align}
    \mathbf{K}^u_{ab}(\mathbf{u}^{(k)}, d_n) \, \Delta\mathbf{u}^{(k)}_b &= -\mathbf{R}^u_a(\mathbf{u}^{(k)}, d_n), \label{eqn:newton_u}\\
    \mathbf{u}^{(k+1)} &= \mathbf{u}^{(k)} + \Delta\mathbf{u}^{(k)}.
\end{align}
With $\mathbf{u}_{n+1}$ fixed, the damage field $d_{n+1}$ is then updated by solving the bound-constrained minimization:
\begin{align}
    \min_{d \in [d_n, 1]} \quad \Pi(\mathbf{u}_{n+1}, d), \label{eqn:damage_min}
\end{align}
which enforces the irreversibility condition $d_{n+1} \geq d_n$. The two subproblems are iterated until the damage increment $\max_a |d^{(k)}_a - d^{(k-1)}_a|$ falls below a prescribed tolerance.

\subsection{Nonlinear solution strategy}\label{sec:4.5}

The discrete residual vectors for the displacement and damage subproblems are obtained by substituting the finite element discretization into the weak forms \eqref{eqn:20} and \eqref{eqn:21}:
\begin{align}
    \mathbf{R}^{u}_{a}
    &= \int_{\Omega_0} (\mathbf{B}^{u}_{a})^{\mathrm{T}}\mathbf{P} \, \mathrm{d}V - \mathbf{f}^{\mathrm{ext}}_a, \label{eqn:28}
    \\
    R^{d}_{a}
    &= \int_{\Omega_0} \left[g'(d)\,\mathcal{H}\,N^{d}_{a}
    + \frac{1}{c_{w}} \left(\frac{1}{2\ell_{d}}\,N^{d}_{a} + \ell_{d} \, (\mathbf{B}^{d}_{a})^{\mathrm{T}} \mathbf{A}\,\nabla d\right) \right] \mathrm{d}V, \label{eqn:29}
\end{align}
where $\mathbf{f}^{\mathrm{ext}}_a = \int_{\Omega_0} N^u_a \mathbf{B}\,\mathrm{d}V + \int_{\partial\Omega_{0,t}} N^u_a \mathbf{T}\,\mathrm{d}A$ is the external force vector.

The consistent tangent matrices are obtained by linearization of the residuals:
\begin{align}
    \mathbf{K}^{u}_{ab}
    &= \frac{\partial \mathbf{R}^u_a}{\partial \mathbf{u}_b} = \int_{\Omega_0} (\mathbf{B}^{u}_{a})^{\mathrm{T}}\mathbb{A}\mathbf{B}^{u}_{b} \, \mathrm{d}V, \label{eqn:30}
    \\
    K^{d}_{ab}
    &= \frac{\partial R^d_a}{\partial d_b} = \int_{\Omega_0} \left[g''(d)\,\mathcal{H}\,N^{d}_{a}N^{d}_{b}
    + \frac{1}{c_{w}}\,\ell_{d} \, (\mathbf{B}^{d}_{a})^{\mathrm{T}} \mathbf{A}\,\mathbf{B}^{d}_{b} \right] \mathrm{d}V, \label{eqn:31}
\end{align}
where $\mathbb{A}=\partial\mathbf{P}/\partial\mathbf{F}$ is the fourth-order elasticity tensor (material tangent modulus).

The discrete residual vectors for each field are assembled into global residual vectors $\mathbf{R}^u \in \mathbb{R}^{n_{\mathrm{dim}} \times n_u}$ and $\mathbf{R}^d \in \mathbb{R}^{n_d}$:
\begin{align}
    \mathbf{R}^u = [R^u_{a,i}], \quad a=1,\ldots,n_u, \quad i=1,\ldots,n_{\mathrm{dim}}, \label{eqn:global_Ru}\\
    \mathbf{R}^d = [R^d_{a}], \quad a=1,\ldots,n_d. \label{eqn:global_Rd}
\end{align}
The global system can be interpreted as the difference between internal and external force vectors:
\begin{align}
    \mathbf{R}^u = \mathbf{F}^u_{\mathrm{int}} - \mathbf{F}^u_{\mathrm{ext}}, \quad
    \mathbf{R}^d = \mathbf{F}^d_{\mathrm{int}}. \label{eqn:R_int_ext}
\end{align}
For the linearization of the residual, the consistent tangent stiffness matrix is assembled from the block matrices defined in Eqs.~\eqref{eqn:30} and \eqref{eqn:31}. In the staggered scheme, each subproblem is solved independently, so the global system decouples into two separate Newton updates:
\begin{align}
    \mathbf{K}^{uu}\,\Delta\mathbf{u} = -\mathbf{R}^u, \quad
    K^{dd}\,\Delta\mathbf{d} = -\mathbf{R}^d, \label{eqn:global_system}
\end{align}
where each system is solved sequentially with the other field held fixed, reflecting the staggered decoupling between the displacement and damage subproblems.

The staggered scheme is summarized in Algorithm~\ref{alg:euclid}. The irreversibility constraint $\dot{d} \geq 0$ is enforced by updating the lower bound $d_n$ at each load step, preventing crack healing.

\begin{center}
  \begin{minipage}{0.95\linewidth}
    \begin{algorithm}[H]
      \caption{Staggered scheme for anisotropic phase-field fracture}\label{alg:euclid}
      \begin{algorithmic}[1]
        \State \textbf{Initialization:} Given $(\mathbf{u}_n, \lambda_{3,n}, d_n)$. Set lower bound $d_{\mathrm{lb}} \leftarrow d_n$.
        \State \textbf{Update boundary conditions:} Apply prescribed loads/displacements at $t_{n+1}$.
        \For{$k = 1, \ldots, k_{\max}$}
        \State \textbf{Displacement step:} Solve Eq.~\eqref{eqn:20} for $\mathbf{u}^{(k)}$ with $d^{(k-1)}$ fixed (SNES solver).
        \State \textbf{Plane stress condensation:} With $\mathbf{u}^{(k)}$ and $d^{(k-1)}$ fixed, solve the plane stress condition $P_{33}=0$ (Eq.~\eqref{eqn:P33}) numerically for $\lambda_3^{(k)}$ via Newton--Raphson iteration (SNES solver).
        \State \textbf{Damage step:} Solve Eq.~\eqref{eqn:21} for $d^{(k)}$ with $(\mathbf{u}^{(k)},\lambda_3^{(k)})$ fixed, subject to $d^{(k)} \geq d_{\mathrm{lb}}$ (TAO solver).
        \State \textbf{Check convergence:} Compute $\epsilon_d = \|d^{(k)} - d^{(k-1)}\|_{\infty}$.
        \If{$\epsilon_d < \mathrm{tol}_{\mathrm{AM}}$}
        \State \textbf{break}
        \EndIf
        \EndFor
        \State \textbf{Update lower bound:} $d_{\mathrm{lb}} \leftarrow d^{(k)}$ (enforce irreversibility).
        \State \textbf{Store solution:} $(\mathbf{u}_{n+1}, \lambda_{3,n+1}, d_{n+1}) \leftarrow (\mathbf{u}^{(k)}, \lambda_3^{(k)}, d^{(k)})$.
      \end{algorithmic}
    \end{algorithm}
  \end{minipage}
\end{center}

\section{Numerical examples}\label{sec:5}

All examples are solved with the alternate minimization scheme described in Algorithm~\ref{alg:euclid} under plane stress. The phase-field length scale is set to the minimum element size, $\ell_d = 2h_{\min}$, determined by locally refined meshes near expected crack paths. The power-law exponents in~\eqref{eqn:anis_H} are $p = q = r = 1$. Reaction forces are obtained by integrating the first Piola--Kirchhoff traction over the loaded boundary and multiplying by the out-of-plane thickness.

\begin{remark}[Numerical regularizations]\label{rmk:numerical_regularization}
Three regularizations are employed to ensure Newton convergence:
\begin{enumerate}[label=(\roman*)]
\item \textbf{Residual stiffness.}
A small parameter $k_{\ell}$ is added to the degradation function, replacing
\begin{align*}
    g(d) \;\longrightarrow\; g(d)+k_{\ell}, \qquad k_{\ell} = 10^{-6},
\end{align*}
to prevent complete loss of ellipticity in fully damaged elements \citep{miehe2010thermodynamically,amor2009regularized}.

\item \textbf{Discriminant shift.}
The spectral decomposition requires the eigenvalues of $\mathbf{C}_{\mathrm{2D}}$ via \eqref{eqn:eigenvalues_C} with discriminant $\Delta = (\mathrm{tr}\,\mathbf{C}_{\mathrm{2D}}/2)^2 - \det\mathbf{C}_{\mathrm{2D}}$. At $\mathbf{F}=\mathbf{I}$, the derivative $1/(2\sqrt{\Delta})$ diverges. This is resolved by the shift
\begin{align*}
    \sqrt{\Delta} \;\longrightarrow\; \sqrt{\Delta + \varepsilon_{\Delta}}, \qquad \varepsilon_{\Delta} = 10^{-12},
\end{align*}
with eigenvalue error $O(\sqrt{\varepsilon_{\Delta}})=O(10^{-6})$.

\item \textbf{Smooth Macaulay bracket.}
The anisotropic energy \eqref{eqn:psi_ani_i} involves $\langle\bar{\mathbf{C}}:\mathbf{a}\otimes\mathbf{a}-1\rangle$, whose kink at $\bar{\mathbf{C}}:\mathbf{a}\otimes\mathbf{a}=1$ causes a tangent discontinuity. It is replaced by the $C^1$-smooth approximation
\begin{align*}
    \langle w \rangle \;\longrightarrow\; \tfrac{1}{2}\!\left(w + \sqrt{w^2 + \varepsilon_s^2}\right), \qquad \varepsilon_s = 10^{-6},
\end{align*}
with approximation error $O(\varepsilon_s)$.
\end{enumerate}
All three regularizations are negligible in magnitude and do not affect the solution once the deformation departs from the initial state.
\end{remark}

\subsection{2D edge-cracked plate under uniaxial tension}

\begin{figure}[!htb]
  \centering  
  \includegraphics[width=0.8\textwidth]{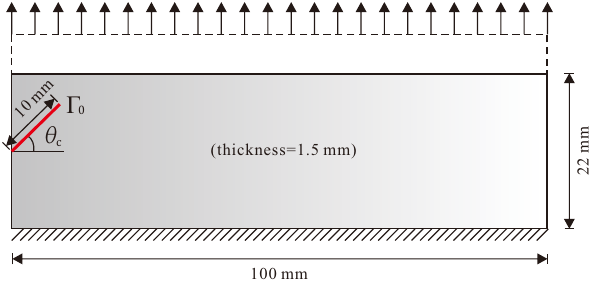}
  \caption{Geometry and boundary conditions of the 2D edge-cracked plate for mixed-mode fracture (\cite{lu2021mixed,pranavi2024unifying}). The initial crack $\Gamma_0$ of length 10~mm is inclined at angle $\theta_c$ measured counter-clockwise from the horizontal ($x$-axis). The bottom boundary is clamped and a prescribed vertical displacement with $u_x = 0$ is applied on the top boundary.}
  \label{fig:MMF_geometry}
\end{figure}

The first example examines a 2D edge-cracked plate under mode~I loading in plane stress. A displacement-controlled approach is applied: the bottom boundary is fully clamped ($\mathbf{u}=\mathbf{0}$), while the top boundary is displaced vertically ($u_x = 0$, $u_y > 0$) via monotonic ramping. For the phase-field, homogeneous Dirichlet conditions $d=0$ are imposed on the top and bottom boundaries to prevent spurious damage initiation at the loaded edges, while the crack is introduced geometrically in the mesh (Figure~\ref{fig:MMF_geometry}).

The material is modeled with the generalized neo-Hookean (GNH) strain energy density and the volumetric--deviatoric--anisotropic decomposition described in Section~\ref{sec:3.2}. The material parameters are calibrated per specimen angle from the pure shear experiments of \citet{lu2021mixed} on Ecoflex~00-30 silicone elastomer (Table~\ref{tab:material_params}). The fracture toughnesses are assumed mode-independent: $G_{c,\mathrm{I}} = G_{c,\mathrm{II}} = G_{c,\mathrm{ani}} = G_c = 0.178$~N/mm (mean of five specimens).

\begin{table}[htbp]
\centering
\caption{Material and phase-field parameters for mixed-mode fracture simulations (Ecoflex 00-30, Lu et al.\ 2021).}
\label{tab:material_params}
\begin{tabular}{ccccccc}
\toprule
$\theta_c$ [deg] & $\mu$ [kPa] & $b$ & $n$ & $\kappa$ [kPa] & $G_{c}$ [N/mm] & $\Gamma_{\mathrm{in}}$ [N/mm] \\
\midrule
0  & 21.4 & 0.0344 & 2 & 19.6 & 0.178 & 0.154 \\
15 & 21.2 & 0.042  & 2 & 19.4 & 0.178 & 0.179 \\
30 & 20.6 & 0.0415 & 2 & 18.9 & 0.178 & 0.156 \\
45 & 20.4 & 0.0347 & 2 & 18.7 & 0.178 & 0.167 \\
60 & 21.0 & 0.04   & 2 & 19.3 & 0.178 & 0.232 \\
\bottomrule
\end{tabular}
\end{table}

\begin{figure}[!htb]
  \centering
  \includegraphics[width=0.9\textwidth]{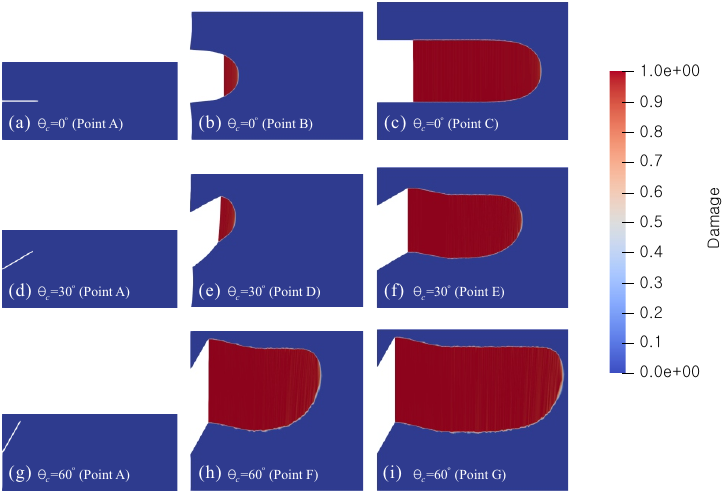}
  \caption{Phase-field damage evolution for the 2D edge-cracked plate at pre-crack angles $\theta_c = 0^\circ$, $30^\circ$, and $60^\circ$. Each row corresponds to a different pre-crack angle, and each column shows the damage field on the current (deformed) configuration at increasing displacement levels (labeled points A--G in Figure~\ref{fig:MMF_force_disp}). Although the initial crack path differs depending on $\theta_c$, all cases eventually converge to horizontal propagation, consistent with the mode~I dominant loading condition.}
  \label{fig:MMF_damage}
\end{figure}

\begin{figure}[!htb]
  \centering
  \includegraphics[width=0.9\textwidth]{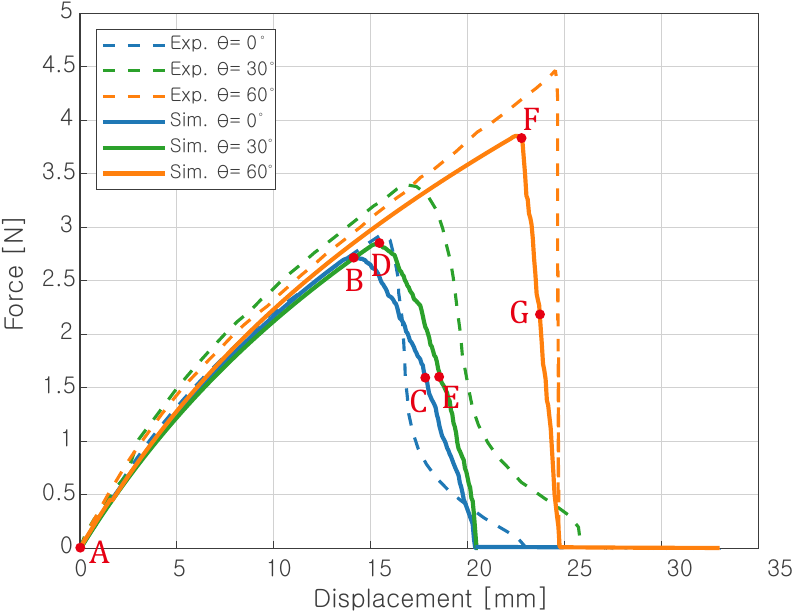}
  \caption{Force--displacement curves for the 2D edge-cracked plate under uniaxial tension at pre-crack angles $\theta = 0^\circ$, $30^\circ$, and $60^\circ$. Dashed lines represent the experimental data of \citet{lu2021mixed}; solid lines represent the present phase-field simulation results. The labeled points (A--G) correspond to the damage snapshots shown in Figure~\ref{fig:MMF_damage}.}
  \label{fig:MMF_force_disp}
\end{figure}

The simulation captures the monotonic ordering of peak loads with $\theta_c$ and the overall shape of the force--displacement response. A systematic under-prediction of the peak force by approximately 10--15\% is observed across all three pre-crack angles (Figure~\ref{fig:MMF_force_disp}). This discrepancy is attributed to the combined effects of the plane-stress assumption, which neglects through-thickness constraint, and uncertainties in the fracture toughness calibration: $G_c = 0.178$~N/mm is taken as the mean of five specimens, while the per-specimen intrinsic toughness $\Gamma_{\mathrm{in}}$ (Table~\ref{tab:material_params}) varies from 0.154 to 0.232~N/mm. The column $\Gamma_{\mathrm{in}}$ reports the experimentally measured intrinsic fracture toughness from \citet{lu2021mixed} and is listed for comparison; the present simulations use the averaged value $G_c$ for all angles.

\subsection{Anisotropic single edge notched specimen under tensile loading}\label{sec:5.2}

\begin{figure}[!htb]
  \centering
  \includegraphics[width=0.5\textwidth]{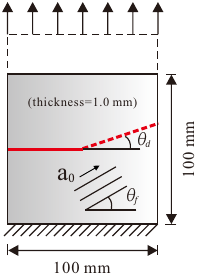}
  \caption{Geometry and boundary conditions of the anisotropic single edge notched (SEN) specimen ($100 \times 100$~mm$^2$, thickness $t = 1.0$~mm). The initial notch extends horizontally to the center of the specimen. The fiber direction $\mathbf{a}_1$ is oriented at angle $\theta_f$ from the horizontal, and $\theta_d$ denotes the crack deflection angle. The bottom boundary is fully clamped and a prescribed vertical displacement is applied on the top boundary.}
  \label{fig:SEN_geometry}
\end{figure}

To isolate and compare the two sources of anisotropy---elastic anisotropy ($\Psi_{\mathrm{ani}}$) and crack density anisotropy ($\mathbf{A}$)---the single edge notched (SEN) specimen is considered \cite{pranavi2024unifying,mandal2020length}. The geometry consists of a square plate with an initial notch extending to the center from the left edge (Figure~\ref{fig:SEN_geometry}). The bottom boundary is fully clamped and the top boundary is displaced vertically via monotonic ramping, with $d = 0$ on both boundaries. The fracture toughness is $G_{c,\mathrm{I}} = G_{c,\mathrm{II}} = G_{c,\mathrm{ani}} = 0.178$~N/mm and $k_2 = 1$ (Section~\ref{sec:3.2}).

\begin{figure}[!htb]
  \centering
  \includegraphics[width=0.8\textwidth]{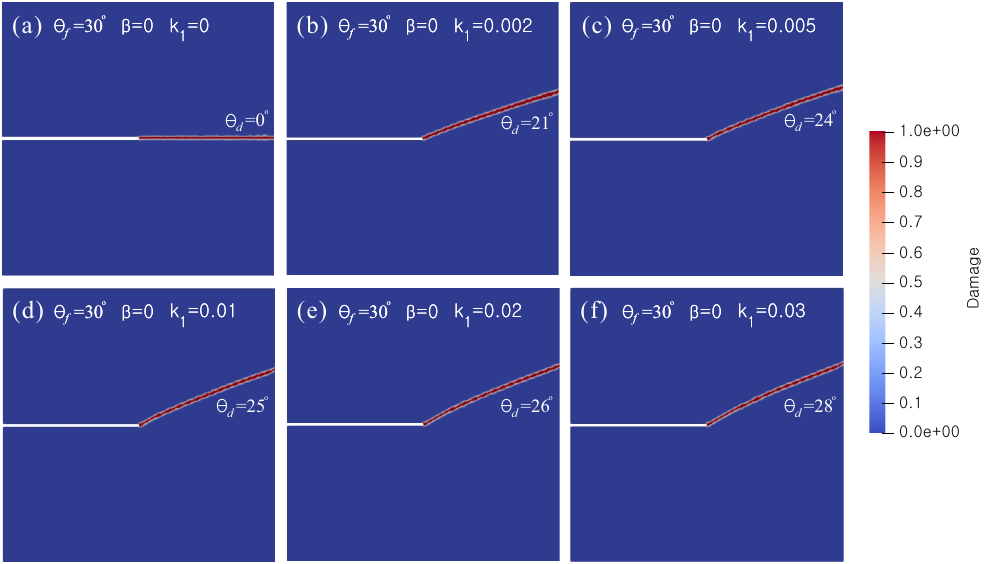}
  \caption{Study~1: Phase-field damage contours for the SEN specimen with pure elastic anisotropy ($\beta = 0$, $\theta_f = 30^\circ$ fixed) and varying fiber stiffness $k_1$. The crack density function is isotropic ($\mathbf{A} = \mathbf{I}$). The deflection angle $\theta_d$ increases rapidly from $0^\circ$ ($k_1 = 0$) to $21^\circ$ ($k_1 = 0.002$) and saturates near $28^\circ$ ($k_1 = 0.03$), approaching but not reaching $\theta_f = 30^\circ$.}
  \label{fig:SEN_study1_damage}
\end{figure}

\begin{figure}[!htb]
  \centering
  \includegraphics[width=0.8\textwidth]{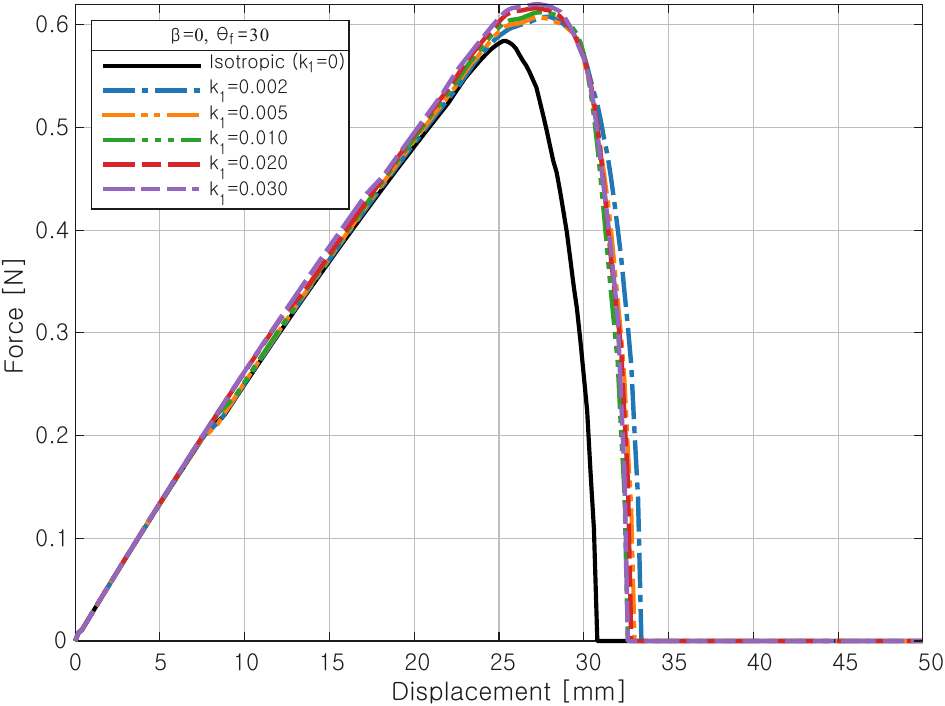}
  \caption{Study~1: Force--displacement curves for the SEN specimen with pure elastic anisotropy ($\beta = 0$, $\theta_f = 30^\circ$ fixed) and varying fiber stiffness $k_1$. The crack density function is isotropic; all directional effects arise solely from the anisotropic strain energy $\Psi_{\mathrm{ani}}$.}
  \label{fig:SEN_study1_k1}
\end{figure}

\begin{figure}[!htb]
  \centering
  \includegraphics[width=0.8\textwidth]{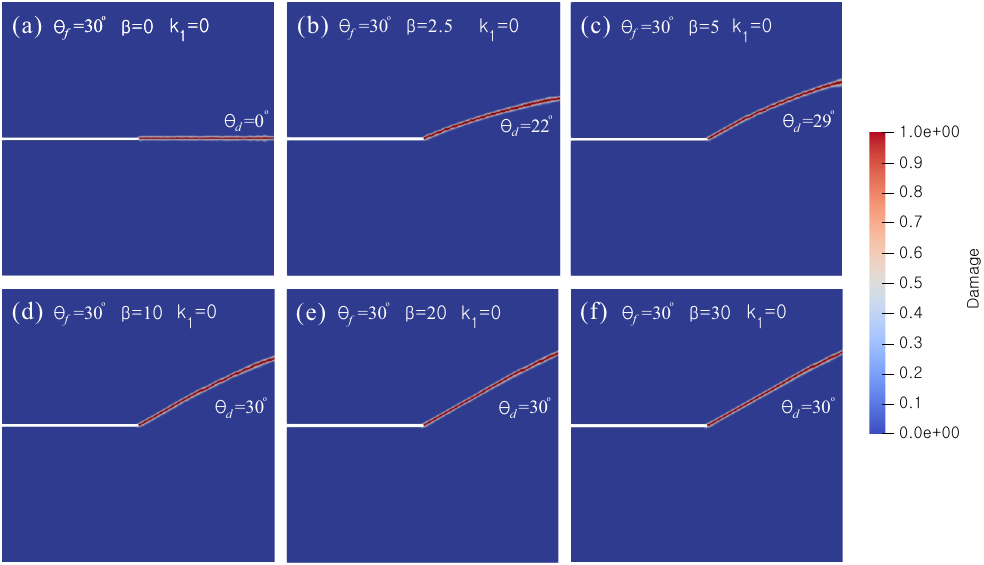}
  \caption{Study~2: Phase-field damage contours for the SEN specimen with pure crack density anisotropy ($k_1 = 0$, $\theta_f = 30^\circ$ fixed) and varying anisotropy strength $\beta$. The crack path progressively deflects toward the fiber direction as $\beta$ increases, demonstrating that the anisotropic crack density function controls crack path deflection.}
  \label{fig:SEN_study2_damage}
\end{figure}

\begin{figure}[!htb]
  \centering
  \includegraphics[width=0.8\textwidth]{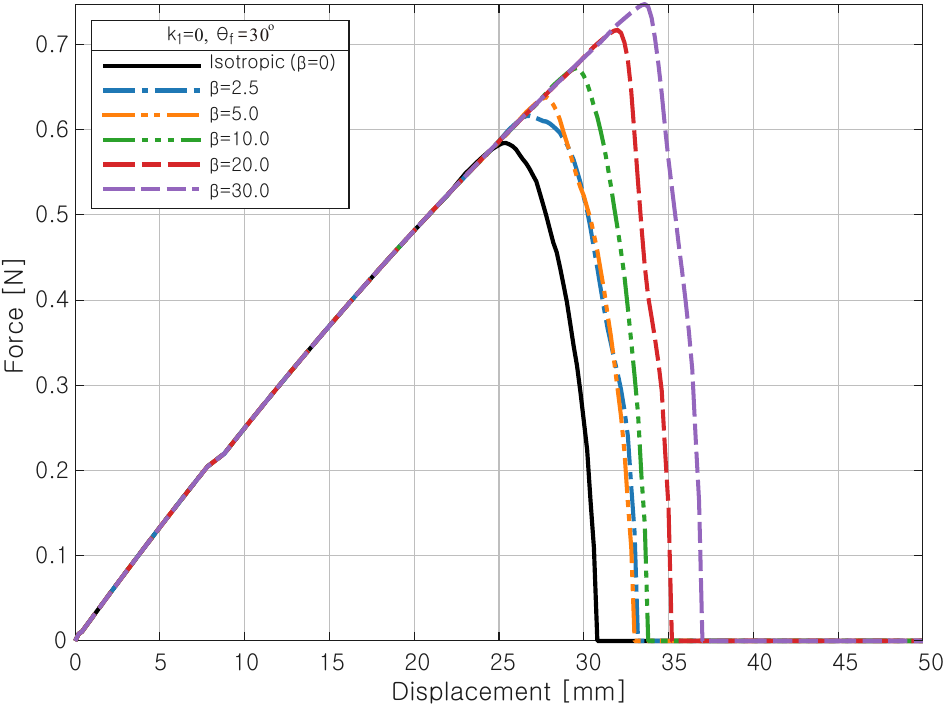}
  \caption{Study~2: Force--displacement curves for the SEN specimen with pure crack density anisotropy ($k_1 = 0$, $\theta_f = 30^\circ$ fixed) and varying anisotropy strength $\beta$. The anisotropic strain energy vanishes; all directional effects arise solely from the anisotropic crack density function $\gamma(\nabla d, \mathbf{A})$.}
  \label{fig:SEN_study2_beta}
\end{figure}

Two parametric studies are designed to isolate the two anisotropy mechanisms:
\begin{itemize}
    \item \textit{Study~1: Pure elastic anisotropy} ($\beta = 0$, $\theta_f = 30^\circ$ fixed). The crack density function is isotropic ($\mathbf{A} = \mathbf{I}$) and the fiber stiffness $k_1$ is varied as $0$, $0.002$, $0.005$, $0.01$, $0.02$, and $0.03$~MPa (Figure~\ref{fig:SEN_study1_damage}). This isolates the effect of the anisotropic strain energy $\Psi_{\mathrm{ani}}$ on the crack driving force (Figure~\ref{fig:SEN_study1_k1}).
    \item \textit{Study~2: Pure crack density anisotropy} ($k_1 = 0$, $\theta_f = 30^\circ$ fixed). The anisotropic strain energy vanishes ($\Psi_{\mathrm{ani}} = 0$) and the anisotropy strength $\beta$ is varied as $0$, $2.5$, $5$, $10$, $20$, and $30$ (Figure~\ref{fig:SEN_study2_damage}). This isolates the effect of the anisotropic crack density function on crack path deflection (Figure~\ref{fig:SEN_study2_beta}).
\end{itemize}

The GNH parameters are $n = 2$, $b = 0.0344$, $\mu = 0.0214$~MPa, and $\kappa = (11/12)\mu$ ($\nu = 0.10$). The phase-field length scale is $\ell_d = 2h_{\min}$ and the power-law exponents are $p = q = r = 1$.

The two parametric studies reveal the distinct and complementary roles of the anisotropic strain energy $\Psi_{\mathrm{ani}}$ and the anisotropic crack density function in governing directional crack propagation. In the following discussion, the fracture toughness is characterized by the area under the force--displacement curve, which represents the total energy absorbed before complete failure.

When the crack density function is kept isotropic ($\beta = 0$, $\mathbf{A} = \mathbf{I}$) and only the fiber stiffness $k_1$ is varied (Figures~\ref{fig:SEN_study1_damage} and~\ref{fig:SEN_study1_k1}), the damage contours reveal that the anisotropic strain energy alone is sufficient to deflect the crack path. The deflection angle $\theta_d$ increases rapidly from $0^\circ$ at $k_1 = 0$ (isotropic) to $21^\circ$ at $k_1 = 0.002$~MPa, then saturates gradually, reaching $28^\circ$ at $k_1 = 0.03$~MPa---close to but not reaching the fiber angle $\theta_f = 30^\circ$. The force--displacement curves show that the initial stiffness remains essentially unchanged across all values of $k_1$, because $\Psi_{\mathrm{ani}}$ only activates once the fiber is stretched beyond unit length. However, an abrupt activation is observed upon transitioning from $k_1 = 0$ to $k_1 = 0.002$~MPa: the peak load rises from approximately 0.58~N at 25~mm to approximately 0.61~N at 28~mm. Beyond $k_1 = 0.002$~MPa, all non-zero $k_1$ curves cluster tightly, with the peak load and fracture displacement remaining nearly identical up to $k_1 = 0.03$~MPa, indicating rapid saturation of both the crack deflection angle and the toughness enhancement. All curves exhibit an abrupt post-peak softening response. The SEN geometry therefore provides insufficient sensitivity to calibrate $k_1$ beyond its binary activation; $k_1$ must instead be identified from experiments on stress-concentration geometries where $\Psi_{\mathrm{ani}}$ modifies the global elastic response (see Section~\ref{sec:5.3}).

When instead the anisotropic strain energy is deactivated ($k_1 = 0$) and only the crack density anisotropy $\beta$ is varied (Figures~\ref{fig:SEN_study2_damage} and~\ref{fig:SEN_study2_beta}), the crack density function likewise deflects the crack, but with a different saturation behavior. The deflection angle increases from $0^\circ$ at $\beta = 0$ to $22^\circ$ at $\beta = 2.5$ and $29^\circ$ at $\beta = 5$, then saturates at $\theta_d = 30^\circ$ (equal to $\theta_f$) for $\beta \ge 10$. Unlike the elastic anisotropy case, the crack density anisotropy achieves the full fiber angle. The force--displacement curves confirm that the initial stiffness remains identical for all values of $\beta$ since the elastic response is purely isotropic. Both the peak load and the displacement at fracture increase monotonically with $\beta$: the isotropic baseline ($\beta = 0$) peaks at approximately 0.54~N near 30~mm, while $\beta = 30$ reaches about 0.75~N at 37~mm. The fracture toughness likewise increases monotonically with $\beta$. The post-peak response transitions from a gradual softening at $\beta = 0$ to a sharp brittle fracture at large $\beta$, reflecting the localization of crack propagation along the fiber-aligned direction.

\subsection{Anisotropic open-hole tension specimen}\label{sec:5.3}

\begin{figure}[!htb]
  \centering
  \includegraphics[width=0.8\textwidth]{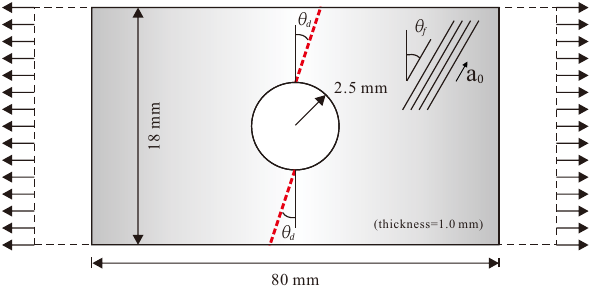}
  \caption{Geometry and boundary conditions of the anisotropic open-hole tension (OHT) specimen ($80 \times 18$~mm$^2$, thickness $t = 1.0$~mm, hole radius $R = 2.5$~mm). Symmetric displacement-controlled tensile loading is applied in the $x$-direction. The fiber angle $\theta_f$ is measured from the $y$-axis: $\theta_f = 0^\circ$ corresponds to the fiber perpendicular to the loading direction, and $\theta_f = 90^\circ$ corresponds to the fiber aligned with loading.}
  \label{fig:OHT_geometry}
\end{figure}

\begin{figure}[!htb]
  \centering
  \includegraphics[width=0.8\textwidth]{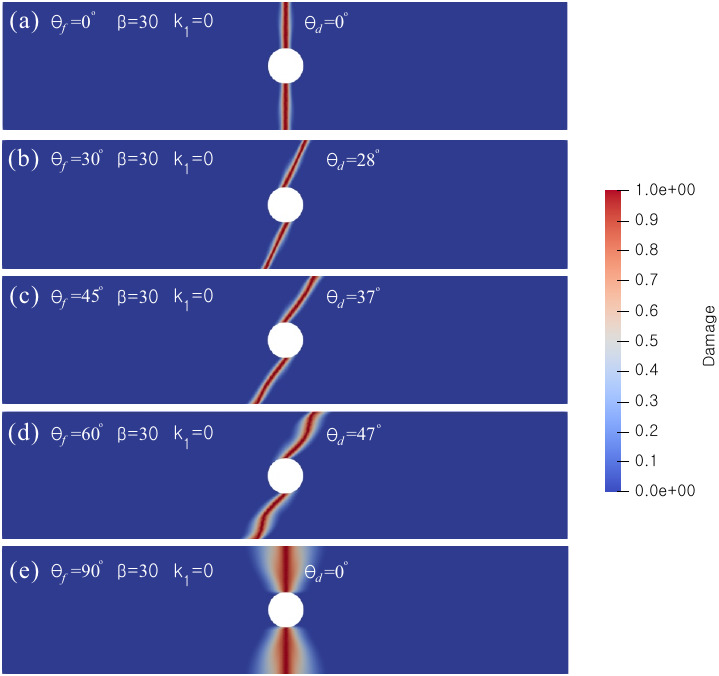}
  \caption{Study~3: OHT damage contours with pure crack density anisotropy ($\beta = 30$, $k_1 = 0$) and varying fiber angle $\theta_f$. The deflection angle $\theta_d$ increases from $0^\circ$ ($\theta_f = 0^\circ$) through $28^\circ$, $37^\circ$, $47^\circ$ ($\theta_f = 30^\circ$--$60^\circ$), and returns to $0^\circ$ at $\theta_f = 90^\circ$ where the fiber is aligned with loading.}
  \label{fig:OHT_study1_damage}
\end{figure}

\begin{figure}[!htb]
  \centering
  \includegraphics[width=0.8\textwidth]{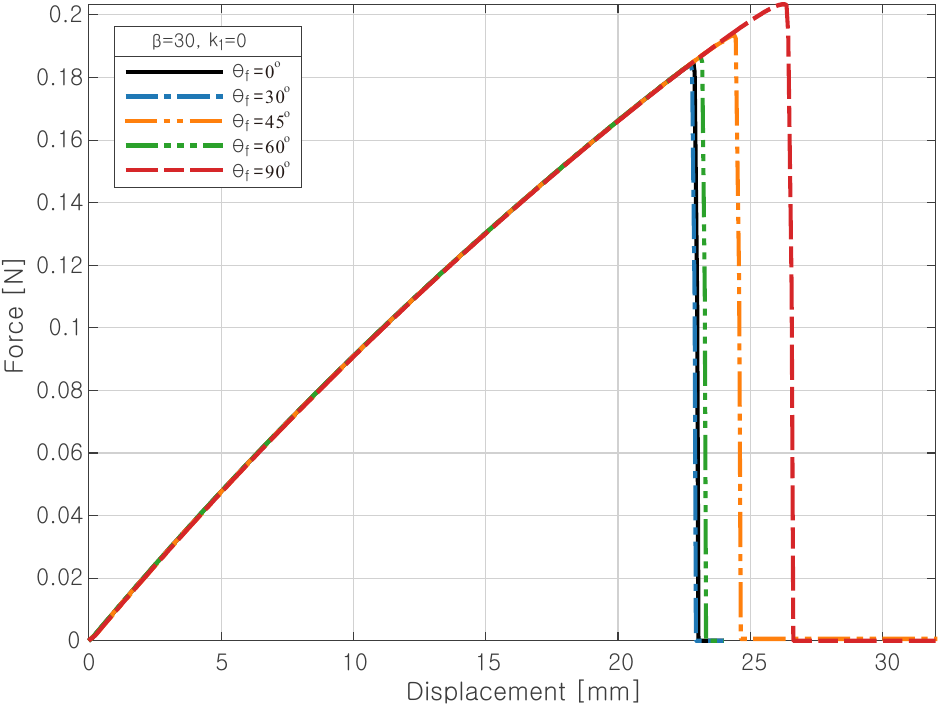}
  \caption{Study~3: OHT force--displacement curves with pure crack density anisotropy ($\beta = 30$, $k_1 = 0$) and varying fiber angle $\theta_f$. The elastic response is identical for all $\theta_f$ since $\Psi_{\mathrm{ani}} = 0$; only $\theta_f = 90^\circ$ shows a distinctly different fracture response.}
  \label{fig:OHT_study1_FD}
\end{figure}

\begin{figure}[!htb]
  \centering
  \includegraphics[width=0.8\textwidth]{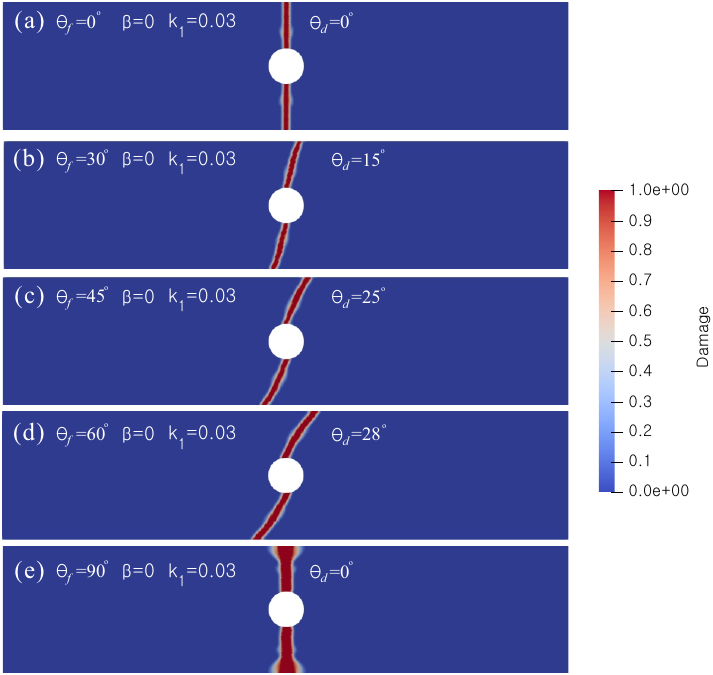}
  \caption{Study~4: OHT damage contours with pure elastic anisotropy ($\beta = 0$, $k_1 = 0.03$~MPa) and varying fiber angle $\theta_f$. The deflection angles ($\theta_d = 0^\circ$, $15^\circ$, $25^\circ$, $28^\circ$, $0^\circ$ for $\theta_f = 0^\circ$--$90^\circ$) are smaller than the crack-density-only case at the same $\theta_f$.}
  \label{fig:OHT_study2_damage}
\end{figure}

\begin{figure}[!htb]
  \centering
  \includegraphics[width=0.8\textwidth]{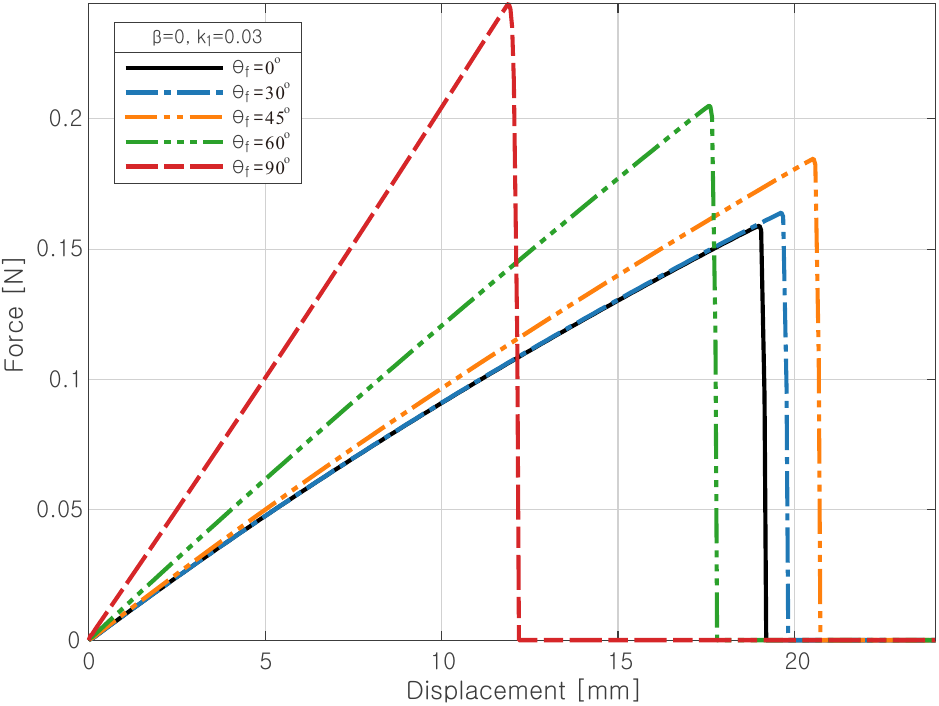}
  \caption{Study~4: OHT force--displacement curves with pure elastic anisotropy ($\beta = 0$, $k_1 = 0.03$~MPa) and varying fiber angle $\theta_f$. The anisotropic strain energy $\Psi_{\mathrm{ani}}$ creates $\theta_f$-dependent initial stiffness, peak force, and fracture displacement---in contrast to the crack-density-only case where these quantities are $\theta_f$-independent.}
  \label{fig:OHT_study2_FD}
\end{figure}

\begin{figure}[!htb]
  \centering
  \includegraphics[width=0.8\textwidth]{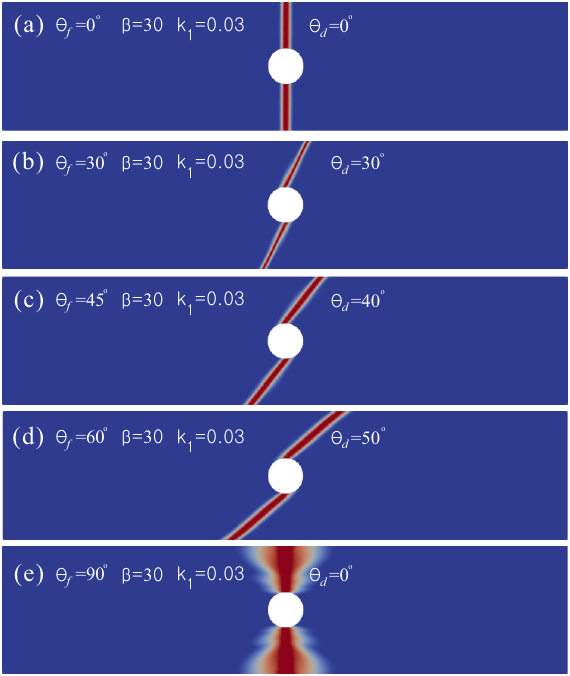}
  \caption{Study~5: OHT damage contours with combined anisotropy ($\beta = 30$, $k_1 = 0.03$~MPa) and varying fiber angle $\theta_f$. The deflection angles ($\theta_d = 0^\circ$, $30^\circ$, $40^\circ$, $50^\circ$, $0^\circ$ for $\theta_f = 0^\circ$--$90^\circ$) exceed those of either mechanism alone, and the crack paths follow the fiber direction more faithfully without the intermediate curvature observed in the individual-mechanism cases.}
  \label{fig:OHT_study3_damage}
\end{figure}

\begin{figure}[!htb]
  \centering
  \includegraphics[width=0.8\textwidth]{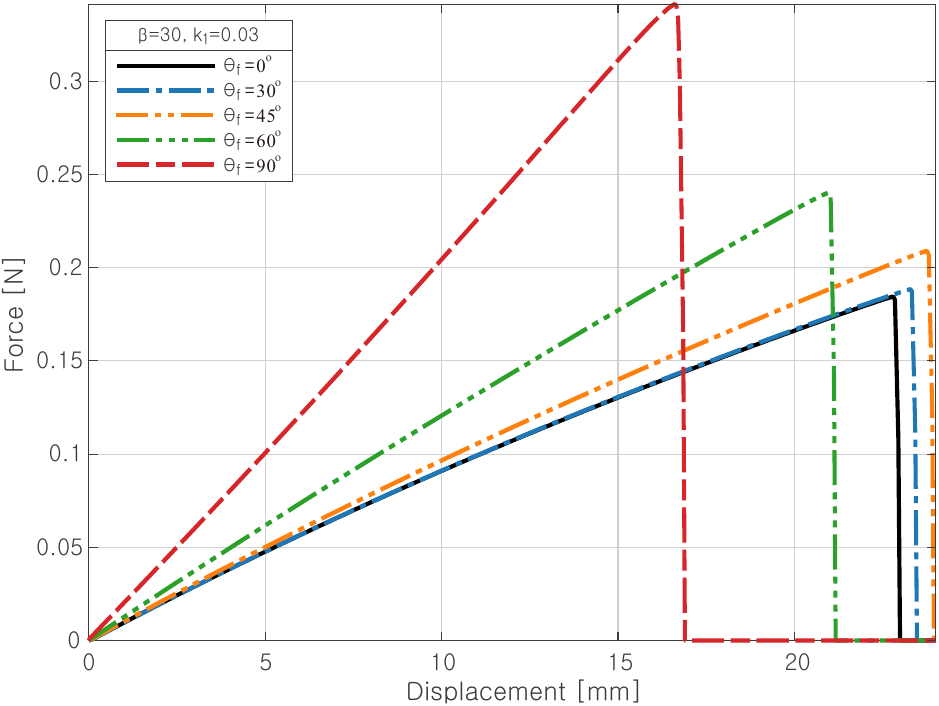}
  \caption{Study~5: OHT force--displacement curves with combined anisotropy ($\beta = 30$, $k_1 = 0.03$~MPa) and varying fiber angle $\theta_f$. The combined effect yields higher peak forces than either mechanism alone, demonstrating nonlinear interaction.}
  \label{fig:OHT_study3_FD}
\end{figure}

To investigate the two anisotropy mechanisms in the presence of stress concentration---where crack nucleation, rather than propagation from a pre-existing notch, governs fracture initiation---an open-hole tension (OHT) specimen is considered. The geometry consists of a rectangular plate ($80 \times 18$~mm$^2$, thickness $t = 1.0$~mm) with a centered circular hole of radius $R = 2.5$~mm (Figure~\ref{fig:OHT_geometry}). Symmetric displacement-controlled tensile loading is applied in the $x$-direction, with $u_y = 0$ on both left and right boundaries and $d = 0$ on both boundaries. The fiber angle $\theta_f$ is measured from the $y$-axis: $\theta_f = 0^\circ$ corresponds to the fiber perpendicular to the loading direction, and $\theta_f = 90^\circ$ corresponds to the fiber aligned with loading. Note that this convention differs from the SEN specimen (Section~\ref{sec:5.2}), where $\theta_f$ is measured from the horizontal ($x$-axis). The fracture toughness is reduced to $G_{c,\mathrm{I}} = G_{c,\mathrm{II}} = G_{c,\mathrm{ani}} = 0.02$~N/mm to ensure crack nucleation at the hole within a feasible strain range, and $k_2 = 1$ (Section~\ref{sec:3.2}). Three parametric studies are performed with the fiber angle $\theta_f$ varied as $0^\circ$, $30^\circ$, $45^\circ$, $60^\circ$, and $90^\circ$:
\begin{itemize}
    \item \textit{Study~3: Pure crack density anisotropy} ($\beta = 30$, $k_1 = 0$). The anisotropic strain energy vanishes ($\Psi_{\mathrm{ani}} = 0$) and the crack density anisotropy deflects the crack path (Figures~\ref{fig:OHT_study1_damage} and~\ref{fig:OHT_study1_FD}).
    \item \textit{Study~4: Pure elastic anisotropy} ($\beta = 0$, $k_1 = 0.03$~MPa). The crack density function is isotropic ($\mathbf{A} = \mathbf{I}$) and the anisotropic strain energy modifies the stress field around the hole (Figures~\ref{fig:OHT_study2_damage} and~\ref{fig:OHT_study2_FD}).
    \item \textit{Study~5: Combined anisotropy} ($\beta = 30$, $k_1 = 0.03$~MPa). Both mechanisms act simultaneously (Figures~\ref{fig:OHT_study3_damage} and~\ref{fig:OHT_study3_FD}).
\end{itemize}

In Study~3 ($\beta = 30$, $k_1 = 0$; Figures~\ref{fig:OHT_study1_damage} and~\ref{fig:OHT_study1_FD}), two contrasting observations emerge from the damage contours and the force--displacement curves. First, the damage contours confirm that the anisotropic crack density function successfully deflects the crack path in the OHT geometry, just as it does in the SEN specimen. The deflection angle $\theta_d$ increases with the fiber angle: $\theta_d = 0^\circ$ at $\theta_f = 0^\circ$ (fiber perpendicular to loading, no preferred deflection direction), rising to $\theta_d \approx 28^\circ$, $37^\circ$, $47^\circ$ at $\theta_f = 30^\circ$, $45^\circ$, $60^\circ$ respectively, and returning to $\theta_d = 0^\circ$ at $\theta_f = 90^\circ$ where the fiber is aligned with the loading direction and the crack propagates symmetrically along the transverse plane.

Second, despite this pronounced variation in crack path geometry, the force--displacement curves reveal that the crack density anisotropy has only a limited effect on the macroscopic mechanical response. Because the anisotropic strain energy is absent ($k_1 = 0$, $\Psi_{\mathrm{ani}} = 0$), the elastic response is purely isotropic: all cases share the same initial stiffness, and the stress field around the hole---hence the crack driving force---is identical regardless of fiber orientation. The curves for $\theta_f = 0^\circ$, $30^\circ$, and $60^\circ$ overlap almost completely, fracturing at $\approx 23$~mm with peak forces of $\approx 0.185$~N. Only $\theta_f = 45^\circ$ is slightly differentiated, fracturing at $\approx 24$~mm. This gradual delay reflects the increasing deflection angle $\theta_d$, which forces the crack to traverse a longer path and thus requires marginally more energy dissipation before complete failure. The $\theta_f = 90^\circ$ case is distinctly different, with a peak force of $\approx 0.20$~N and fracture delayed to $\approx 26$~mm, because the structural tensor $\mathbf{A} = \mathbf{I} + \beta\,\mathbf{a}_1\otimes\mathbf{a}_1$ with the fiber along the loading direction ($x$-axis) maximally enhances the fracture resistance perpendicular to the predominant crack propagation direction ($y$-axis), requiring substantially more energy to nucleate and propagate the crack.

This limited sensitivity of the global force--displacement response to fiber orientation---despite pronounced differences in crack path---highlights a key characteristic of the crack density anisotropy when used alone in stress-concentration geometries: it primarily controls where the crack goes, while its influence on the macroscopic mechanical response remains secondary, arising only indirectly through the crack path length rather than through the elastic strain energy distribution.

A notable observation in Figure~\ref{fig:OHT_study1_damage}(e) is that the $\theta_f = 90^\circ$ case exhibits a wide, diffuse damage band rather than a sharp, localized crack. This is an artifact of the classical structural tensor approach: $\mathbf{A} = \mathbf{I} + \beta\,\mathbf{a}\otimes\mathbf{a}$ with the fiber along the loading direction penalizes the damage gradient component along the fiber but leaves the perpendicular component (the crack propagation direction) with the baseline penalty, resulting in an orientation-dependent damage band width. This artifact is also observed in the elastic-anisotropy-only case (Figure~\ref{fig:OHT_study2_damage}(e)) and the combined case (Figure~\ref{fig:OHT_study3_damage}(e)), indicating that it arises independently from both mechanisms when the fiber is aligned with loading. The quantitative peak force values reported for $\theta_f = 90^\circ$ should therefore be interpreted with the understanding that this artifact contributes to the computed energy dissipation. A consistent formulation in which both the potential and gradient terms of the crack surface density carry the anisotropic resistance---as proposed by \citet{prajapati2026physically}---would eliminate this orientation-dependent band width.

In Study~4 ($\beta = 0$, $k_1 = 0.03$~MPa; Figures~\ref{fig:OHT_study2_damage} and~\ref{fig:OHT_study2_FD}), the results are strikingly different. The damage contours show that $\Psi_{\mathrm{ani}}$ alone also deflects the crack path, but to a lesser extent than the crack density anisotropy: the deflection angle reaches $\theta_d \approx 15^\circ$, $25^\circ$, $28^\circ$ at $\theta_f = 30^\circ$, $45^\circ$, $60^\circ$ respectively, compared to $28^\circ$, $37^\circ$, $47^\circ$ in the crack-density-only case (Study~3). This systematic reduction is consistent with the SEN observation that the elastic anisotropy produces less crack path deflection than the crack density anisotropy (Section~\ref{sec:5.2}). As in Study~3, $\theta_d = 0^\circ$ at both $\theta_f = 0^\circ$ and $90^\circ$ by symmetry.

The force--displacement curves, however, reveal a clear differentiation across all fiber angles: the initial stiffness increases monotonically with $\theta_f$, reflecting the progressively increasing contribution of $\Psi_{\mathrm{ani}}$ to the elastic energy. The fracture displacement varies substantially with $\theta_f$: $\theta_f = 90^\circ$ fractures earliest ($\approx 12$~mm) while $\theta_f = 0^\circ$ fractures latest ($\approx 18$~mm), because the stiffer response at higher $\theta_f$ drives the local energy past the fracture threshold at a smaller applied displacement. The peak forces increase monotonically with $\theta_f$, from $\approx 0.16$~N at $\theta_f = 0^\circ$ to $\approx 0.24$~N at $\theta_f = 90^\circ$. This is the key distinction from the SEN geometry: in the SEN specimen, the elastic anisotropy deflects the crack but the force--displacement response remains nearly unchanged across $k_1$ values; in the OHT specimen, $\Psi_{\mathrm{ani}}$ modifies the entire stress field around the hole, producing $\theta_f$-dependent initial stiffness, fracture displacement, and peak force. It should come as no surprise that the anisotropic driving force plays a significantly expanded role in stress-concentration geometries, since the local strain energy distribution around the hole governs crack nucleation.

Comparing the damage contours across Studies~3--5 (Figures~\ref{fig:OHT_study1_damage},~\ref{fig:OHT_study2_damage}, and~\ref{fig:OHT_study3_damage}) reveals that the combined anisotropy produces crack paths that follow the fiber direction more faithfully than either mechanism alone. In the crack-density-only case (Figure~\ref{fig:OHT_study1_damage}), the deflection angles are $\theta_d \approx 28^\circ$, $37^\circ$, $47^\circ$ at $\theta_f = 30^\circ$, $45^\circ$, $60^\circ$; in the elastic-anisotropy-only case (Figure~\ref{fig:OHT_study2_damage}), they are smaller at $\theta_d \approx 15^\circ$, $25^\circ$, $28^\circ$. When both mechanisms act together (Figure~\ref{fig:OHT_study3_damage}), the deflection angles increase to $\theta_d \approx 30^\circ$, $40^\circ$, $50^\circ$---exceeding either individual mechanism---and the crack paths are straighter, following the fiber orientation without the intermediate curvature observed in the individual-mechanism cases. This indicates that the two mechanisms reinforce each other not only in the force--displacement response but also in directing the crack path geometry.

The force--displacement curves (Figure~\ref{fig:OHT_study3_FD}) exhibit a $\theta_f$-dependent stiffness pattern similar to the elastic-anisotropy-only case, but with substantially higher peak forces. The nonlinear synergy is most evident when comparing the peak force gap between $\theta_f = 0^\circ$ and $90^\circ$ across the three studies. In Study~3 (crack density anisotropy alone), this gap is approximately $0.015$~N ($0.20$ vs $0.185$~N); in Study~4 (elastic anisotropy alone), it is approximately $0.08$~N ($0.24$ vs $0.16$~N). When both mechanisms act together, the gap widens to approximately $0.16$~N ($0.34$ vs $0.18$~N)---roughly 1.6 times the sum of the individual gaps ($0.095$~N). This nonlinear interaction confirms that the two mechanisms interact synergistically: the crack density anisotropy elevates the fracture resistance along the deflected path while $\Psi_{\mathrm{ani}}$ simultaneously raises the stored elastic energy available to drive the crack, and the coupling of enhanced resistance and enhanced driving force amplifies the total energy required for fracture beyond the sum of the individual contributions. Note that the $\theta_f = 90^\circ$ peak values include contributions from the orientation-dependent damage band width artifact discussed above, so the reported synergy should be regarded as an upper bound.

\section{Conclusions}\label{sec:6}

This paper presents a thermodynamically consistent phase-field formulation for anisotropic mixed-mode fracture in finite deformation hyperelasticity. A normalized mixed-mode crack driving force is formulated as an additive combination of volumetric, deviatoric, and anisotropic contributions, each normalized by the corresponding fracture toughness and governed by tunable power-law exponents ($p$, $q$, $r$). Combined with the anisotropic crack density function $\gamma(\nabla d,\mathbf{A})$, the formulation provides two independent and complementary mechanisms for modeling directional fracture. The framework is first validated against the mixed-mode fracture experiments of \citet{lu2021mixed} on an edge-cracked soft elastomer plate, where the simulated force--displacement responses and crack paths show good agreement with experimental data across three pre-crack angles. Systematic parametric studies on single-edge-notched (SEN) and open-hole tension (OHT) specimens with a single fiber family then elucidate the distinct physical roles of these two mechanisms. The crack density anisotropy $\gamma(\nabla d,\mathbf{A})$ primarily controls the crack path direction, deflecting it toward the fiber orientation, while its influence on the macroscopic force--displacement response remains secondary. Conversely, the anisotropic strain energy $\Psi_{\mathrm{ani}}$ also deflects the crack but with rapid saturation in $k_1$; its role depends critically on the geometry. In the SEN specimen the force--displacement response is insensitive to $k_1$, whereas in the OHT specimen $\Psi_{\mathrm{ani}}$ modifies the entire stress field around the hole, producing $\theta_f$-dependent stiffness, peak force, and fracture displacement---making the OHT geometry essential for calibrating $k_1$. When both mechanisms act simultaneously, the combined effect exceeds the sum of individual contributions, confirming a nonlinear synergistic interaction. These results indicate that fibers oriented at an angle to the crack create a material-induced mode-mixity that cannot be captured by the crack density anisotropy alone, necessitating the inclusion of $\Psi_{\mathrm{ani}}$ in the driving force.

The proposed framework is not without its limitations. The current implementation is restricted to quasi-static loading without inertia effects. The near-incompressible limit ($\kappa \gg \mu$) requires careful treatment of the volumetric penalty to avoid locking. The present AT1 model employs a simple linear geometric function $\alpha(d)=d$ with a quadratic degradation function, which makes the predicted peak load and energy dissipation sensitive to the length-scale parameter $\ell_d$ and, consequently, to the mesh size. Adopting a rational degradation function whose parameters depend on material elasticity and fracture properties---as proposed by \citet{mandal2020length} following the framework of \citet{wu2017unified}---would render the results length-scale insensitive. Furthermore, the present formulation applies the same softening behavior to the isotropic and anisotropic parts of the strain energy, whereas physically the matrix and fiber constituents may exhibit distinct softening characteristics; incorporating separate degradation functions for the isotropic and anisotropic contributions would improve the physical fidelity of the model. The classical structural tensor approach introduces anisotropy only through the gradient term of the crack surface density function, while the potential term retains the isotropic fracture toughness; a consistent formulation in which both terms carry the anisotropic resistance symmetrically---as in \citet{prajapati2026physically}---would eliminate the orientation-dependent damage band width artifact, which was observed in the OHT simulations at $\theta_f = 90^\circ$ (Figures~\ref{fig:OHT_study1_damage}(e),~\ref{fig:OHT_study2_damage}(e), and~\ref{fig:OHT_study3_damage}(e)) and contributes to the computed peak forces at that fiber orientation. Extension to multiple fiber families with independent anisotropy parameters would enable application to bidirectional composites and woven laminates. Systematic validation against experimental mixed-mode fracture data---particularly Arcan-type tests \citep{arcan1978method} and anisotropic crack kinking experiments---would strengthen confidence in the calibration of the interaction exponents.

The two anisotropy mechanisms in phase-field fracture play fundamentally different and geometry-dependent roles. The crack density anisotropy governs the crack path (fracture resistance), while $\Psi_{\mathrm{ani}}$ governs the crack driving force; in stress-concentration geometries, $\Psi_{\mathrm{ani}}$ additionally controls the elastic strain energy distribution around the stress concentrator. Both mechanisms are indispensable, and their interaction is nonlinear. This understanding provides a clear physical basis for selecting and calibrating anisotropy parameters in phase-field models of fiber-reinforced composites.

\section*{CRediT authorship contribution statement}
\textbf{Guk Heon Kim:} Data curation, Writing -- original draft. \textbf{Minseo Kim:} Visualization, Writing -- original draft. \textbf{Kwangsan Chun:} Writing -- review \& editing, Funding acquisition. \textbf{Jaemin Kim:} Methodology, Software, Validation, Formal analysis, Investigation, Data curation, Visualization, Writing -- original draft, Writing -- review \& editing, Project administration, Supervision.

\section*{Acknowledgments}
\begin{itemize}
    \item This research is funded by the `Changwon National University - Samsung Changwon Hospital joint Collaboration Research Support Project, South Korea' in 2025.
    \item This study is conducted as part of the Glocal University Project, supported by the RISE (Regional Innovation System \& Education) program funded by the Ministry of Education.
\end{itemize}

\section*{Declarations}
The authors declare no competing financial interests or personal relationships that could have appeared to influence the work reported in this paper.

\begin{appendices}

\section{Derivation of crack evolution equation and irreversibility}
\label{sec:A}
\renewcommand{\thefigure}{A\arabic{figure}}
\setcounter{figure}{0}
\renewcommand{\theequation}{A\arabic{equation}}
\setcounter{equation}{0}

Multiplying the test function $\delta d$ with the strong form of damage evolution (Equation \ref{eqn:7}):
\begin{equation}\label{eqn:A1}
    \int_{\Omega_0} \left(\omega-\nabla\cdot\boldsymbol{\xi}\right)\delta d\,\mathrm{d}\Omega_0 = 0
\end{equation}
Applying the product rule and divergence theorem,
\begin{align}\label{eqn:A2}
    \int_{\Omega_0} \omega\,\delta d\,\mathrm{d}\Omega_0
    + \int_{\Omega_0} \boldsymbol{\xi}\cdot\nabla\delta d\,\mathrm{d}\Omega_0
    - \int_{\partial\Omega_0} \left(\boldsymbol{\xi}\cdot\mathbf{N}\right)\delta d\,\mathrm{d}S
    = 0.
\end{align}
With the natural boundary condition $\boldsymbol{\xi}\cdot\mathbf{N}=0$ on $\partial\Omega_{0}\setminus\Gamma_0$, it follows that
\begin{align}\label{eqn:A3}
    \int_{\Omega_0} \omega\,\delta d\,\mathrm{d}\Omega_0
    + \int_{\Omega_0} \boldsymbol{\xi}\cdot\nabla\delta d\,\mathrm{d}\Omega_0
    = 0.
\end{align}
Substituting the constitutive relations---where $\Psi^{+} = G_{c}\mathcal{H}$ with $\mathcal{H}$ the normalized mixed-mode driving force (cf.\ Section~\ref{sec:2.8})---and the AT1 functional ($\alpha(d)=d$, $g(d)=(1-d)^{2}$) used in the implementation, and noting that $G_{c}$ factors out of every $d$-dependent term and cancels ($G_{c}>0$), yields
\begin{align}\label{eqn:A4}
    \int_{\Omega_0} \left[g'(d)\,\mathcal{H}\,\delta d
    + \frac{1}{c_{w}} \left(\frac{1}{2\ell_{d}}\,\delta d + \ell_{d} \nabla\delta d\cdot\mathbf{A}\,\nabla d\right) \right]\,\mathrm{d}\Omega_0 = 0.
\end{align}
The strong form of the crack evolution equation can be written as
\begin{align}
    \mathnormal{f}:= g'(d)\,\mathcal{H}
    + \frac{1}{c_{w}} \left(\frac{1}{2\ell_{d}} - \ell_{d} \nabla\cdot\left(\mathbf{A}\,\nabla d\right)\right) = 0 \quad \text{in} \quad \Omega_0. \label{eqn:A5}
\end{align}
Taking together the crack irreversibility condition and dissipation inequality, the formulation can be written in terms of the Karush--Kuhn--Tucker (KKT) conditions \citep{karush1939minima,kuhn1951nonlinear}:
\begin{align}
    \dot{d} \geq 0
    ,\quad
    \mathnormal{f} \geq 0
    ,\quad
    \dot{d}\mathnormal{f} = 0
    \quad &\text{in} \quad \Omega_0. \label{eqn:A6}
\end{align}
These equations correspond to the weak form of the crack evolution equation (Equation \ref{eqn:21}).
  

\end{appendices}

\bibliographystyle{elsarticle-harv} 
\bibliography{references}

\end{document}